\documentclass[preprint2]{aastex}

\usepackage{amssymb}
\usepackage{amsmath}

\newcommand{\axpeinstein}{1E\,1547.0-5408}
\newcommand{\axp}{1E\,1841-045}

\newcommand{\axprxs}{1RXS\, J1708\,-\,4009}
\newcommand{\axpu}{4U\,0142\,$+$\,61}

\newcommand{\gr}{$\gamma$-rays}

\slugcomment{}

\shorttitle{X- and \gr\ characteristics of \axpeinstein}
\shortauthors{Kuiper et al.}

\begin{document}



\title{Temporal and spectral evolution in X- and $\gamma$-rays of magnetar \axpeinstein\ since
its October 2008 outburst: the discovery of a transient hard pulsed component after its January 2009 outburst}

\author{L. Kuiper\altaffilmark{1}, W. Hermsen\altaffilmark{1,2}, P.R. den 
Hartog\altaffilmark{3} and J.O. Urama\altaffilmark{4}}
\email{L.M.Kuiper@sron.nl}
\altaffiltext{1}{SRON-National Institute for Space Research, Sorbonnelaan 2, 
3584 CA, Utrecht, The Netherlands }
\altaffiltext{2}{Astronomical Institute ``Anton Pannekoek", University of 
Amsterdam, Science Park 904, 1098 XH Amsterdam, The Netherlands}
\altaffiltext{3}{Stanford University HEPL/KIPAC Physics, 382 via Pueblo Mall Stanford, 94305, USA}
\altaffiltext{4}{Dept. of Physics \& Astronomy, University of Nigeria, Nsukka, Nigeria}


\begin{abstract}
The magnetar \axpeinstein\ exhibited outbursts in October 2008 and January 2009. In this paper we present in great detail
the evolution of the temporal and spectral characteristics of the persistent {\it total} and {\it pulsed} emission of \axpeinstein\ 
between $\sim$1 and 300 keV starting in October 3, 2008, and ending in January 2011. We analyzed data collected with the {\it Rossi X-ray Timing Explorer}, the {\it International Gamma-Ray Astrophysics Laboratory} and the {\it Swift} satellite. 
We report the evolution of the pulse frequency, and the measurement at the time of the onset of the January 2009 outburst of an insignificant jump in frequency, but a major frequency derivative jump $\Delta{\dot\nu}$ of $+(1.30\pm0.14)\cdot 10^{-11}$ Hz/s ($\Delta\dot{\nu}/\dot{\nu}$ of $-0.69\pm0.07$). Before this $\dot{\nu}$ glitch, a single broad pulse is detected, mainly for energies below $\sim$ 10 keV. Surprisingly, $\sim$ 11 days after the glitch a new transient high-energy (up to $\sim$ 150 keV) pulse appears with a Gaussian shape and width 0.23, shifted in phase by $\sim$ 0.31 compared to the low-energy pulse, which smoothly fades to undetectable levels in $\sim$ 350 days. We report the evolution of the {\it pulsed} emission spectra. For energies 2.5--10 keV all {\it pulsed} spectra are very soft with photon indices $\Gamma$ between -4.6 and -3.9. For $\sim$ 10--150 keV, after the $\dot{\nu}$ glitch, we report hard non-thermal {\it pulsed} spectra, similar to what has been reported for the persistent pulsed emission of some Anomalous X-ray Pulsars. This pulsed hard X-ray emission reached maximal luminosity 70$\pm$30 days after the glitch epoch, followed by a gradual decrease by more than a factor 10 over $\sim$300 days. These characteristics differ from those of the {\it total} emission. Both, the {\it total} soft X-ray (1--10 keV) and hard X-ray (10--150 keV) fluxes were maximal already two days after the 2009 January outburst, and decayed by a factor of $\gtrsim$3 over $\sim$400 days.   
The {\it total} spectra can be described with a black-body (kT values varying in the range 0.57--0.74 keV) plus a single power-law model. The photon index exhibited a hardening ($\sim$-1.4 to $\sim$-0.9) with time, correlated with a decrease in flux in the 20--300 keV band. We discuss these findings in the framework of the magnetar model.
\end{abstract}

\keywords{pulsars: individual (\axpeinstein; SGR J1550-5418; \axprxs; \axpu; \axp), X-rays: stars}


\section{Introduction}
Over the last decade, observational evidence has mounted that Anomalous X-ray pulsars (AXPs) and Soft Gamma-ray Repeaters (SGRs) belong to the same class of objects. Most publications discuss these objects in the frame work of strongly magnetized neutron stars, dubbed magnetars \citep{duncan1992, thompson1995}. Alternative interpretations exist including `normal' magnetized neutron stars with fall-back disks \citep[e.g.][]{chatterjee2000, alpar2001}, white dwarfs \citep[e.g.][]{malheiro2011}, or even quark stars \citep[e.g.][]{xu2007, orsaria2011}. In this paper we adopt the magnetar interpretation. Magnetars are rotating neutron stars with surface magnetic field strengths of $10^{14} - 10^{15}$ G, well above the critical field strength of $4.413 \times 10^{13}$ G at which the cyclotron energy of an electron reaches the electron rest mass energy. They have long rotation periods $P$ (in the range 2--12 s) and large period derivatives $\dot{P}$ ($\sim 10^{-11}$ s s$^{-1}$). 

Their soft X-ray ($<$10 keV) luminosities of $\sim 10^{35}$ erg s$^{-1}$ exceed the available rotational energy losses \citep[see reviews by][]{woods2006,kaspi2007,mereghetti2008}. In addition to SGR giant flares, SGRs and AXPs exhibit periods of high activity wherein radiative outbursts are accompanied by numerous short bursts of typical durations a few hundred milliseconds. In the context of the magnetar twisted magnetosphere model \citep{thompson2002}, the energy release is due to magnetic field re-arrangement which may be triggered by crustal deformation causing a glitch in their rotational timing behaviour.

A distinct feature is the high non-thermal luminosity of the persistent emission of AXPs above 20 keV, first esthablished by \citet{kuiper04} with the discovery of pulsed hard X-ray emission up to 150 keV from AXP 1E 1841-045 in Supernova remnant (SNR) Kes 73. The X-ray luminosity spectrum exhibits two peaks, one near 1 keV and the other above 100 keV \citep[see e.g.][]{kuiper06,gotz2006,denhartog2008a,denhartog2008b,enoto2010b}.
The soft X-ray part of the spectrum below 10 keV and the hard X-ray part above 20 keV, both exhibit luminosities exceeding the spin-down power 1-3 orders of magnitude. The spectra of the first component can be empirically described by a thermal black body (BB) plus a soft power-law (index 2-4) model or two BB models, and more physically with resonant cyclotron scattering models \citep{thompson2002, lyutikov2006, fernandez2007, guever2007, guever2008, rea2008, nobili2008, zane2009, zane2011}.

On the origin of the hard non-thermal persistent component above $\sim 10$ keV, however, no concensus has been reached. It has been shown \citep{denhartog2008a, denhartog2008b} that this component is persistent and within the statistical errors of about 20\% stable over years, possibly as long as a decade, in both total and pulsed fluxes and spectral shapes, as well as in pulse phase/shape. Most recent attempts to explain these findings in the magnetar interpretation are by \citet[][]{beloborodov2007,beloborodov2009,baring2007,baring2008,pavan2009}. The non-thermal component above 10 keV has also been discussed in the context of the fall-back disk model \citep{truemper2010a,truemper2010b}.
In his latest work, \citet{beloborodov2009,beloborodov2011} discusses how a ``starquake'' can cause convective motions in the crust which twist the magnetic field anchored to the surface, after which it gradually untwists, dissipating magnetic energy and producing radiation. A current carrying bundle of closed field lines (``j-bundle'') is created above the magnetic dipole axis. Part of the thermal X-rays emitted by the neutron star are upscattered by the inner relativistic outflow in the quasi-steady j-bundle, producing a beam of hard X-rays. 

In this context, it is therefore of great interest to study the evolution of the observational characteristics of a magnetar before and after a glitch followed by a radiative outburst; to study the spectral variations in the persistent steady and pulsed emissions in the soft and hard X-ray bands, as well as variations in pulse profile. Sofar the decay from a radiative outburst could only be studied in detail for energies below $\sim$ 10 keV, e.g. for 1E2259+586 \citep{woods04}, SGR 0501+4516 \citep{rea2009, gogus2010a}, SGR J1833-0832 \citep{gogus2010b}, SGR 1900+14 \citep{gogus2011}, SGR1833-0832 \citep{esposito2011} and CXOU J164710.2-455216 \citep{woods2011}. In this work we present the first study in which both the temporal and spectral characteristics at both soft and hard X-ray energies before and during a radiative outburst of a magnetar, \axpeinstein, have been derived.

The X-ray source \axpeinstein\ was discovered by \citet{lamb1981} in an Einstein HRI observation.  Based on its X-ray spectrum and variability, \citet{gelfand07} discussed this source as candidate magnetar in a candidate supernova remnant (G327.24-0.13). Subsequently, its period was discovered in the radio band by \citet{camilo07}, making it the first magnetar of which the pulsation was not discovered at X-ray energies, and the second known radio-emitting magnetar after the transient XTE J1810-197 \citep{camilo2006}. With the reported period of 2.069 s
it appeared to be the fastest spinning magnetar known, with a derived surface magnetic dipole field strength of $2.2 \times 10^{14}$ G, a characteristic age of 1.4 kyr, and a spin-down luminosity of  $1.0 \times 10^{35}$ ergs s$^{-1}$, assuming a distance of $\sim$9 kpc (estimated from the dispersion measure). X-ray pulsations were first detected with XMM-Newton by \citet{halpern08} in a period of enhanced activity in 2007. These authors concluded that the source was recovering from an X-ray outburst in between 2006 and 2007.

On Oct. 3, 2008 (MJD 54742) \axpeinstein\ started a period of strong bursting activity detected by Swift \citep{krimm2008a,
krimm2008b, israel2010}, and by the Gamma-ray Burst Monitor (GBM) on the Fermi Gamma-ray Space Telescope
\citep{kaneko10}. By the latter authors the source was dubbed SGR J1550-5418, motivated by the hundreds of typical 
soft-gamma-repeater bursts. The bursting activity was accompanied by a radiative outburst which was also detected with 
RXTE and Chandra \citep{ng2011}. \citet{israel2010} used Swift and \citet{ng2011} RXTE and Chandra data to study the evolution of the pulse profile shape and the phase-averaged spectrum for energies below 10 keV during the outbursting activity.

The Swift Burst Alert Telescope (BAT) reported renewed extreme bursting activity starting on Jan. 22, 2009 (MJD 54853) \citep{gronwall09}, also detected by INTEGRAL \citep{savchenko2009, mereghetti2009a}, the Fermi GBM \citep{connaughton2009}, Konus-Wind \citep{golenetskii2009}, RHESSI \citep{bellm2009}, and the Wide-band All-sky Monitor aboard Suzaku \citep{terada2009}. Follow-up observations with Swift and XMM-Newton revealed dust scattering X-ray rings centered on \axpeinstein, from which \citet{tiengo10} deduced a source distance of $\sim$3.9 kpc. Based on a 100 ks public Target-of-Opportunity (ToO) INTEGRAL observation \citet{baldovin2009} reported the discovery of a hard power-law tail in the spectrum.
Exploiting also part of our 600 ks INTEGRAL open-time (PI den Hartog) ToO observations, a power-law index of $\sim$ 1.5 was found for energies from 20 up to 150 keV from 2 to 7 days after the Jan. 2009 outburst \citep{denhartog2009}. This was confirmed in Suzaku observations taken 7 days after the outburst \citep{enoto2010a}. Furthermore, \citet{kuiper2009a} announced the detection with INTEGRAL of pulsed emission for energies up to 150 keV.

\citet{enoto2010a} presented the Suzaku results on AXP \axpeinstein\ for their observations 7 days after the onset of the bursting activity in 2009. Pulsations were detected for energies up to 70 keV, and the total, time-averaged 0.7-114 keV spectrum was discussed. \citet{ng2011} compared the 2008 and 2009 outbursts, analyzing RXTE and Chandra observations taken during $\sim$ 20 days after the onset of the bursting activities in 2008 and 2009, comparing pulse profiles, pulsed-emission spectra below 10 keV, and spin evolution. Furthermore, \citet{bernardini2011} monitored the 2009 outburst with Chandra, XMM-Newton and INTEGRAL over a period of two weeks after the onset, as well as with Swift over a 1.5 year interval. Finally, \citet{scholz2011} focussed on the 2009 outburst using Swift XRT observations for energies below 10 keV addressing the persistent radiative evolution and a statistical study of the burst properties.

In this work we exploited the extensive data base of observations of \axpeinstein\ with RXTE, INTEGRAL and Swift covering the 27 months from the onset of the bursting activities in October 2008 till January 2011. We studied in detail the variable (total and pulsed) emission, but did not address the burst properties \citep[see e.g.][]{savchenko2010, mereghetti09, kaneko10}. 
The evolution of the timing parameters, pulse profiles, fluxes, spectra have been derived over the broad energy range $\sim$ 1 - 300 keV. Particularly interesting is the discovery of a transient non-thermal hard X-ray component, which appeared shortly after the onset of Jan. 22, 2009 outburst, triggered by a $\dot{\nu}$ timing glitch, as a distinct new pulse in the pulse profile.
Finally, our findings are summarized in Sect. \ref{sec:sum} and are compared with model predictions within the magnetar framework.


\section{Instruments and observations}

\subsection{Rossi X-ray Timing Explorer}

In this study extensive use is made of data from monitoring observations
of AXPs with the two non-imaging X-ray instruments aboard RXTE, the Proportional Counter Array 
(PCA; 2-60 keV) and the High Energy X-ray Timing Experiment (HEXTE; 15-250 keV). 

\subsubsection{RXTE PCA}
\label{pca_char}
The PCA \citep{jahoda96} consists of five collimated xenon proportional 
counter units (PCUs) with a total effective area of $\sim 6500$ cm$^2$ over a $\sim 1\degr$ 
(FWHM) field of view. Each PCU has a front Propane anti-coincidence layer and three Xenon 
layers which provide the basic scientific data, and is sensitive to 
photons with energies in the range 2-60 keV. The energy resolution is about 18\% at 6 keV.
All PCA data used in this study have been collected from observations in {\tt GoodXenon} mode 
allowing high-time resolution ($0.9\mu$s) analyses in 256 spectral bins.
Since the launch of RXTE on Dec. 30, 1995 the PCA has experienced high voltage breakdowns for 
all constituting PCUs at irregular times. To avoid further breakdowns, during its already 14.5 
year mission not all PCUs are simultaneously operating. The most stable PCU is PCU-2, which is
on for almost all of the time. On average one (50\%) or two (40\%) PCUs is/are operational during
a typical observation.

\subsubsection{RXTE HEXTE}
\label{hexte_char}
The HEXTE instrument \citep{rothschild98} consists of two independent detector 
clusters A\& B, each containing four Na(Tl)/ CsI(Na) scintillation
detectors. The HEXTE detectors are mechanically collimated to a $\sim 1\degr$ (FWHM) 
field of view and cover the 15-250 keV energy range with an energy resolution of 
$\sim$ 15\% at 60 keV. The collecting area is 1400 cm$^2$ taking into account the 
loss of the spectral capabilities of one of the detectors. The maximum time 
resolution of the tagged events is $7.6\mu$s. In its default operation mode the 
field of view of each cluster is switched on and off source to provide instantaneous 
background measurements. However, also HEXTE suffers from aging, and since July 13, 2006
HEXTE cluster-A operates in staring mode at an on-source position, while
on March 29, 2010 cluster-B was commanded to stare at an off-source position.

Due to the co-alignment of HEXTE and the PCA, they simultaneously observe celestial targets. 
Table \ref{obs_table} lists the RXTE observations used in this study with in the fourth column the screened 
 exposure of PCU-2 (see Sect \ref{rxte_evol}). 
A typical observation consists of several sub-observations spaced more or less uniformly (4---9 days apart)
between the start and end date of the observation cycle. However, for a $\sim 2-3$ weeks time period 
directly after the two outbursts intensive timing has been performed. 
\begin{table}[t]
\caption{List of RXTE observations of \axpeinstein\ used in this study. \label{obs_table}}
{\footnotesize
\begin{center}
\begin{tabular}{cccc}
\hline
\textbf{Observation} & \multicolumn{2}{c}{\textbf{Begin/End Date}}    & \textbf{Exp.$^{[2]}$}\\
\textbf{identifier}  & \multicolumn{2}{c}{\textbf{(yyyy/mm/dd)}}      & \textbf{(ks)}\\
\hline
20060            & 1997-05-15          & 1997-05-15           &    0.376$^{[1]}$\\
\hline
93017            & 2008-10-03          & 2009-01-24           &  162.484\\
94017            & 2009-01-25          & 2009-06-22           &  174.676\\
94427            & 2009-06-30          & 2009-12-26           &  104.677\\
95312            & 2010-01-04          & 2010-12-25           &  236.952\\
93017 -- 95312   & 2008-10-03          & 2010-12-25           &  678.789\\
\hline\hline
\multicolumn{4}{l}{$^{[1]}$Scanning observation} \\
\multicolumn{4}{l}{$^{[2]}$PCU-2 exposure after screening} \\
\end{tabular}
\end{center}}
\end{table}
\subsection{INTEGRAL}
\label{instr_integral}
The INTEGRAL spacecraft \citep{winkler03}, launched 17 October 2002, carries two main 
$\gamma$-ray instruments: a high-angular-resolution imager IBIS \citep{ubertini03} and
a high-energy-resolution spectrometer SPI \citep{vedrenne03}. The payload is further
supported by two monitor instruments providing complementary observations in the X-ray 
and optical energy bands, the Joint European Monitor for X-rays \citep[JEM-X; ][]{lund2003} and 
the Optical Monitoring Camera (OMC; 500 - 600 nm (Johnson V-filter)).  All the high-energy 
instruments make use of coded aperture masks enabling image reconstruction in the hard 
X-ray/soft $\gamma$-ray band.

In our study, guided by sensitivity considerations, we only used data recorded by the INTEGRAL 
Soft Gamma-Ray Imager ISGRI \citep{lebrun03}, the upper detector system of IBIS, sensitive 
to photons with energies in the range $\sim$20 keV -- 1 MeV, and JEM-X operating in the 3-35 keV 
X-ray band.
 
With an angular resolution of about $12\arcmin$ and a source location accuracy of better than 
$1\arcmin$ (for a $>10\sigma$ source) ISGRI is able to locate and separate high-energy sources 
in crowded fields within its $19\degr \times 19\degr$ field of view (50\% partially coded) with 
an unprecedented sensitivity ($\sim$ 960 cm$^2$ at 50 keV).
Its energy resolution of about 7\% at 100 keV is amply sufficient to determine the (continuum) spectral 
properties of hard X-ray sources in the $\sim$ 20 - 300 keV energy band.

The timing accuracy of the ISGRI time stamps recorded on board is about $61\mu$s. The time 
alignment between INTEGRAL and RXTE is better than $\sim 50\mu$s, verified using data 
from simultaneous RXTE and INTEGRAL observations of the accretion-powered millisecond pulsars 
IGR J00291+5934 \citep{falanga05} and IGR J17511-3057 \citep{falanga11}.

JEM-X consists of two identical telescopes each having a field of view of $7\fdg5$ (diameter) at half response 
and able to pinpoint a $15\sigma$ source with a 90\% location accuracy of about $1\arcmin$. 
Its energy resolution and timing accuracy ($3\sigma$) are 1.3 keV at 10 keV and $122.1\mu$s, respectively. 
\begin{table}[t]
\caption{List of INTEGRAL observations of \axpeinstein\ used in this work, sorted on INTEGRAL orbital revolutions. In the last two columns are given the exposures and the time segment identifiers. \label{obsint_table}}
{\footnotesize
\begin{center}
\setlength{\tabcolsep}{2pt}
\begin{tabular}{ccccc}
\hline
\textbf{Revs.}      & \textbf{Begin/End Date} & \textbf{Begin/End} &\textbf{Exp.$^{[1]}$} &\textbf{id.} \\
                    & \textbf{(yyyy/mm/dd)}   & \textbf{MJD}       &\textbf{(ks)}         &             \\
\hline\hline
731                 & 2008-10-08           / 2008-10-10           & 54747.9 -- 54749.2 & \ 98.4  & 1 \\
767                 & 2009-01-24           / 2009-01-25           & 54855.6 -- 54856.9 & \ 92.6  & 2 \\
768-772             & 2009-01-28           / 2009-02-08           & 54859.6 -- 54870.7 &  575.6  & 3 \\
782-791             & 2009-03-09           / 2009-04-08           & 54899.6 -- 54929.0 & 191.9  & 4 \\
840-850             & 2009-08-30           / 2009-10-01           & 55073.6 -- 55105.5 & 407.8  & 5 \\
899-910             & 2010-02-23           / 2010-03-29           & 55250.3 -- 55284.3 & 524.0  & 6 \\
911-912             & 2010-03-30           / 2010-04-03           & 55285.9 -- 55289.0 & 193.7  & 7 \\
\hline
\hline
\multicolumn{5}{l}{$^{[1]}$Effective on-axis exposure} \\
\end{tabular}
\end{center}}
\end{table}

In its default operation mode INTEGRAL observes the sky in a dither pattern 
with $2\degr$ steps, which could be rectangular e.g. a $5 \times 5$ dither pattern 
with 25 grid points, or hexagonal with 7 grid points (target in the middle). Typical integration
times for each grid point (pointing/sub-observation) are in the range 1800 - 
3600 seconds. This strategy drives the structure of the INTEGRAL data archive which is 
organised in so-called science windows (Scw) per INTEGRAL orbital revolution (lasting for 
about 3 days) containing the data from all instruments for a given pointing. 
Most of the INTEGRAL data reduction in this study was performed with the Offline 
Scientific Analysis (OSA) version 7.0 distributed by the INTEGRAL Science Data Centre 
\citep[ISDC; see e.g.][]{courvoisier03}.

Table \ref{obsint_table} lists the INTEGRAL orbital revolution (Rev.) identifiers with corresponding start/end dates of the
observations used in the imaging/spectral analyses and/or timing analyses of \axpeinstein.
Revolutions 731 (Oct. 2008; ToO), 767 (Jan. 2009; public ToO), 768--772 (Jan-Feb., 2009; ToO) and 911--912 had 
\axpeinstein\ as on-axis prime target, while during revolutions 782--791, 840--845/850 and 899--910 (all part of 
an INTEGRAL key program to observe deeply the {\it l\/}=$\pm25$ Galactic plane region) \axpeinstein\ was in the field 
of view at moderate off-axis angles.

\subsection{Swift}
\label{instr_swift}
The {\it Swift} satellite \citep{gehrels04} was launched on November 20, 2004, and regular observations began on April 5, 2005.
The main goal of the mission is the study of gamma-ray bursts and their afterglows. {\it Swift\/} carries three
co-aligned instruments: The wide-field coded aperture mask Burst Alert Telescope (BAT; 15-150 keV), the narrow field 
($23\farcm 6 \times 23\farcm 6$) grazing incidence Wolter 1 X-Ray Telescope (XRT; 0.2-10 keV) and 
the Ultraviolet/Optical Telescope (UVOT). 

During periods of non-burst operations observations of other high-energy sources are scheduled.
In this work we used data of \axpeinstein\ gathered by the XRT \citep{burrows05}. 

Regular monitoring of \axpeinstein\ commenced
on June 22, 2007 following the detection of radio pulsations on June 8, 2007 \citep{camilo07}.
The XRT has several operation modes of which we used only the Windowed Timing (WT) and Photon-counting (PC) modes.
In PC mode full imaging ($600 \times 600$ pixels; pixel scale $2\farcs 36$/pixel) and spectroscopic resolution are retained, 
but its time resolution is only 2.5073 s, insufficiently to detect the pulsations of \axpeinstein. In WT mode 10 rows are compressed
in one, and only the central 200 columns are read out. Therefore only a $\sim 8\arcmin$ central strip of the field of view is covered.
The time resolution in WT mode, however, is 1.7675 ms amply sufficient for pulse timing studies of \axpeinstein.

Intensive monitoring data are available for 2--3 weeks time periods directly after the onsets of the Oct. 2008 and Jan. 2009 
outbursts. We used the XRT WT data gathered between Oct. 3, 2008 and Oct. 24, 2008 (Swift observation ids. 00330353000---00330353016;
MJD 54742--54763) for a combined RXTE PCA/Swift XRT timing study of the pulsed signal at soft X-rays.

From the onset of the Jan. 2009 outburst on Jan. 22, 2009 till Feb. 7, 2009 \axpeinstein\ was observed daily by the XRT both in WT and
PC mode. This period of dense sampling was followed by monitoring observations taken at a rate of roughly 1--4 times per month in either
WT or PC mode, and this strategy continues up to now.
XRT WT data gathered between Jan. 22, 2009 and Feb. 22, 2009 (MJD 54853--54884) have been used in the construction of timing models
in combination with RXTE PCA data.

Data taken in PC mode on Jan. 25 \& 29 and Feb. 4, 2009 (Swift observation ids. 00341114000, 00030956034 and 00030956039, respectively) covering the time window for which we have deep INTEGRAL observations (see Table \ref{obsint_table}; Jan. 24 - Feb. 8, 2009 i.e. MJD 54855.6-54870.7), have been employed in spectral analyses of the {\em total} soft X-ray emission of \axpeinstein.
Furthermore, to study the evolution of the total soft X-ray spectrum we analyzed Swift XRT observations 00030956048 (July 8, 2009), 00030956051 (Aug. 19, 2009), 00090404003 (Apr. 12, 2010), 00090404019 (Sept. 28, 2010) and 00090404027 (Feb. 25, 2011).

Finally, XRT observations in WT mode performed between Jan. 4 and 13, 2009 and after Feb. 22, 2009 have been used for verification purposes of timing models based on merely RXTE (monitoring) observations. The latter are sampled too sparsely given the high level of timing noise \citep{camilo08} present in this source.

\section{Timing, total flux and pulsed flux evolutions below $\sim$ 10 keV}
\label{rxte_evol}

We exploited the regular monitoring observations with the PCA on RXTE over more than 2 years (see Table \ref{obs_table}) that provided the required high-statistics data. Because the number of active PCUs at any time was changing, we treated the five PCUs constituting the PCA separately.
Good time intervals (GTI) have been determined for each PCU by including only time periods when the PCU in question is on, 
and during which the pointing direction is within $0\fdg 05$ from the target, the elevation angle above Earth's horizon 
is greater than $5\degr$, a time delay of 30 minutes since the peak of a South-Atlantic-Anomaly passage holds, and a 
low background level due to contaminating electrons, as measured by PCU-2, is observed.
Furthermore, periods during which a detector break-down (see Sect.\ref{pca_char}) has occurred, are excluded for further analysis.
Finally, because we are only interested in the persistent non-burst emission properties, we rejected small duration ($\le 1$ s) 
bursts. The latter filtering is especially important just after the onset of the Oct. 2008 and Jan. 2009 outbursts, when copious
numbers of bursts are detected.
The resulting good time intervals have subsequently been applied in the screening process to the data streams from each of 
the PCUs (e.g. see Table \ref{obs_table} for the resulting screened exposure of PCU-2 per observation cycle).

The TT (Terrestrial Time) arrival times of the selected events (for each sub-observation and PCU unit) were converted to 
arrival times at the solar system barycenter (in Barycentric Dynamical Time (TDB) time scale) using 1) the JPL DE200 solar system ephemeris, 2) the instantaneous spacecraft position, and 3) the sub-arcsecond celestial position of \axpeinstein.
The position used is:  $(\alpha,\delta)=(15^{\hbox{\scriptsize h}} 50^{\hbox{\scriptsize m}}  54\fs 11,-54\degr 18\arcmin 23\farcs 7)$ 
for epoch J2000 \citep{camilo07}, which corresponds to (l,b)=(327.23705,-0.13162) in Galactic coordinates. 

\begin{figure*}[t]
\centerline{\includegraphics[height=17cm,width=14cm,angle=0,bb=10 85 570 765,clip=]{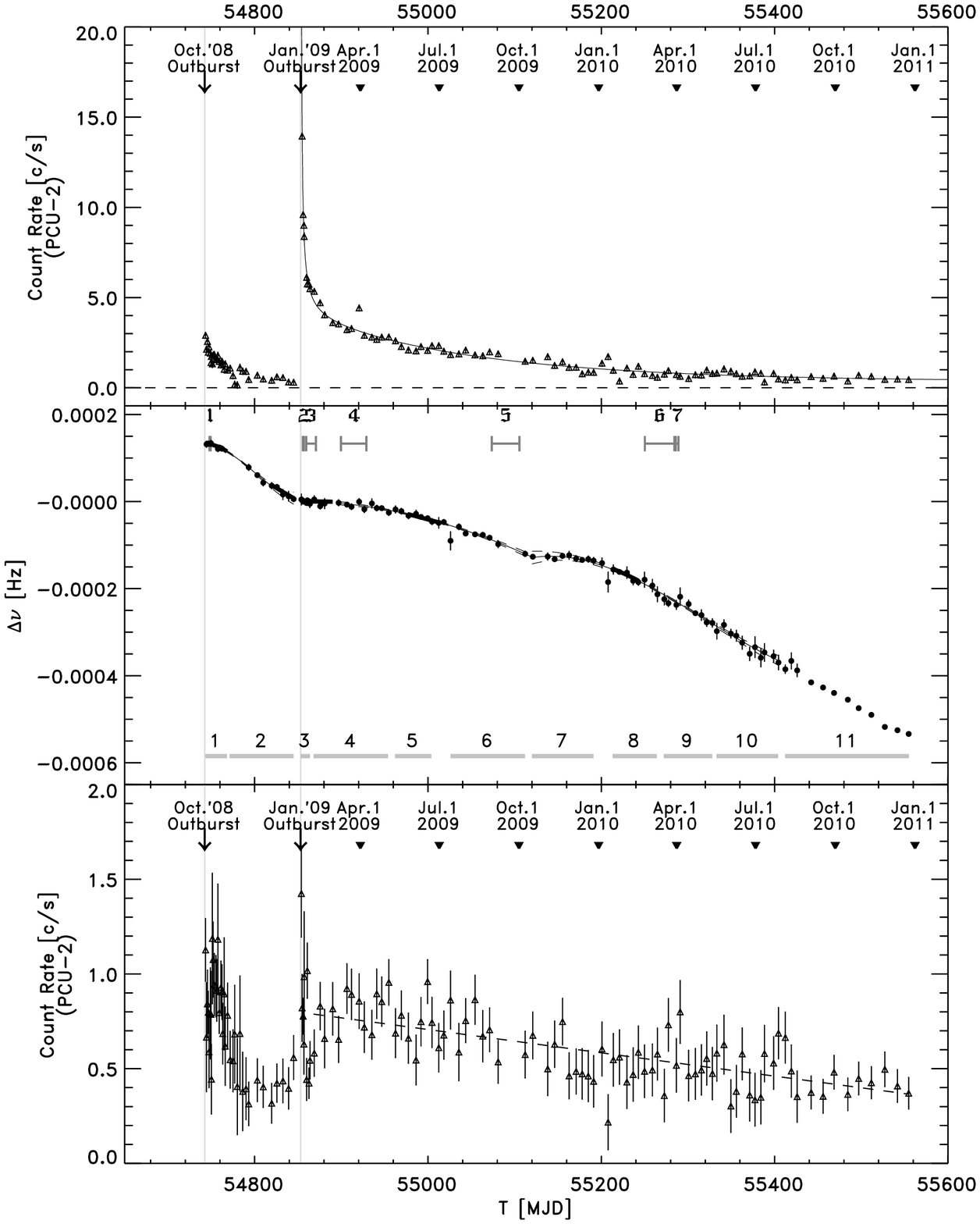}}
  \caption{\label{evolution_flux_nu_pflux} The evolution of the total flux (top panel) and pulsed flux (bottom panel)
  of \axpeinstein\ from Oct. 3, 2008 till December 25, 2010 as measured by the PCA/PCU-2, summing the signals from all three 
  detection layers for the pulse-height-channel range 4--27 ($\sim 2-10$ keV). We have subtracted the 1997 pre-outburst reference level (see section \ref{rxte_evol_total} ) for
  the total rate (top panel; dashed line) and superposed the best fit decay model as a solid line (top panel). 
  The middle panel shows the frequency measurements (data points), three incoherent fit segments (solid plus
  dashed lines ($=1\sigma$ uncertainty)) and five bold segments representing reliable coherent timing solutions.
  Time intervals for which detailed PCA pulse morphology and spectral studies
  were performed are labeled at the bottom of the middle panel (11 segments). At the top of the middle panel 7 intervals
  are shown for which INTEGRAL observations are available (see Table \ref{obsint_table}). The dashed line in the bottom panel represents the 
  best fit of a linear decay model for the period MJD 54868-55555.}
\end{figure*}

\subsection{Total flux evolution at X-rays below $\sim$ 10 keV}
\label{rxte_evol_total}

From the screened event datasets and the break-down/burst corrected exposure times we could derive the total count rate from a
circular field with a radius of $\sim 1\degr$ (PCA field of view up to zero collimator response) on \axpeinstein. 
We determined this for the 4--27 PHA channel range ($\sim 2-10$ keV) including all Xenon detectors layers.
The count rate includes the instrumental and celestial (Galactic diffuse) background, emission from \axpeinstein\ 
and its PWN/SNR \citep{vink09} and other discrete point sources. We found in the archive two RXTE scanning observations, both 
performed on May 15, 1997, with \axpeinstein\ in the field of view for 40 s (screened) at $28\farcm 1$ offset angle and 336 s (screened) at $17\farcm 3$ offset angle. We used the count rate from the combination of these 2 scanning observations as reference, because the 
source at that time was very likely in quiescence.

The total flux evolution, from Oct. 3, 2008 up to December 25, 2010, of \axpeinstein\ above the reference level of $9.70\pm 0.16$ c/s 
as measured by PCU-2 in the 4--27 PHA range (all detector layers) is shown in the upper panel of Fig. \ref{evolution_flux_nu_pflux}.

We cropped this figure in vertical direction in order to better visualize the late time evolution (the count rate of 44.19(8) measured
directly after the Jan. 22, 2009 outburst at MJD 54853.911 is off scale).
The onsets of the Oct. 3, 2008 and Jan. 22, 2009 outbursts are indicated in this figure. 
There is a monitoring gap of about a month which occurred in Sept./Oct. 2009 due to RXTE spacecraft anomalies.

It is clear that since the Oct. 3, 2008 the total flux decayed to almost its reference level just before the Jan. 22, 2009 outburst.
The flux decay of the much more intense second outburst, however, has still not reached its reference level $\sim 2$ year after the onset.
We studied the decay of the total flux since the Jan. 22, 2009 outburst in more detail by fitting the data, excluding the time intervals in which the mini-outbursts occurred (see Sect.\ref{sect:mini_outbursts}), with a model composed of a constant, an exponential- and a
power-law like decay component: 
\begin{equation}
R_{\hbox{\scriptsize{tot}}}(t) = a + b \cdot \left(1+\frac{t-t_0}{\tau_0}\right)^{\alpha} + c \cdot \exp(-\frac{t-t_0}{\tau_1})
\label{eq:xray_decay}
\end{equation}
We fixed $t_0$ at MJD$=54853.035$, used only times with $t\ge 54854$ (MJD) and fitted the 6 free parameters to obtain the best desciption of the observed decay.
We found the following optimized parameters: $a= 0.33 \pm 0.05$ c/s, $b= 152.5 \pm 4.4$ c/s, $c = 3.5 \pm 0.1$ c/s, $\tau_0=0.129 \pm 0.004$ day, $\tau_1=207 \pm 11$ day, and finally $\alpha=-1.01 \pm 0.01$ (errors are $1\sigma$), for a fit with a reasonable quality of $\chi_{\nu}^{2}=138.58/82=1.69$. This model is superposed in the top panel of Fig. \ref{evolution_flux_nu_pflux} as a solid line, and follows the measurements globally pretty well. The slowly decaying exponential component has a time scale of typically $\sim 200$ days, whereas the rapidly decaying power-law component contributes only significantly to the total flux for the first $\sim 50$ days since the Jan. 22, 2009 outburst, over which its fractional contribution to the total rate is more than 10\%. The power-law component dominates only during a time period of $\sim 5$ days directly after the outburst, then the combination of constant plus exponential component takes over.

\begin{figure}[ht]
\centerline{\includegraphics[height=6cm,width=7.5cm,angle=0,bb=40 155 565 660,clip=]{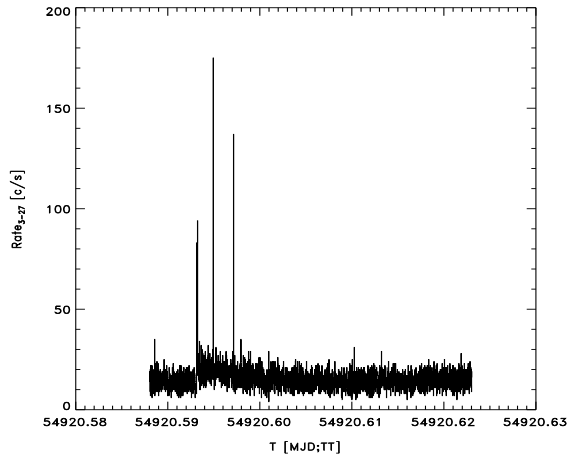}}
  \caption{\label{mini_outburst_one_rate} Count rate of \axpeinstein\ in the 3--27 PHA range (all detector layers combined) as
     observed by the PCU-2 during observation 94017-09-09-00 on Mar. 30, 2009 (MJD 54920). The time bin width is 1 s.
     The time is expressed in MJD for the TT time system. 
     Note the enhanced emission level of $\sim 450$ s duration with superposed three short bursts.}
\end{figure}

\begin{figure}[ht]
  \begin{center}
    \includegraphics[height=6cm,width=7.5cm,angle=0,bb=40 155 565 658,clip=]{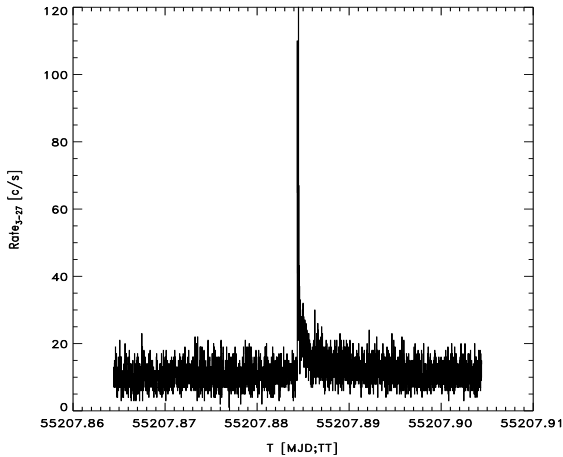}
    \caption{\label{mini_outburst_two_a} Count rate of \axpeinstein\ for the observation 95312-01-02-00 on Jan. 11, 2010 (MJD 55207) as in 
                                         Fig. \ref{mini_outburst_one_rate}. Note the enhanced emission level of $\sim 130$ s duration with
                                         superposed two bursts separated at 11 s (not resolved at the displayed scale).}
  \end{center}
\end{figure}
\begin{figure}[ht]
  \begin{center}
    \includegraphics[height=6cm,width=6.5cm,angle=0,bb=60 160 540 660,clip=]{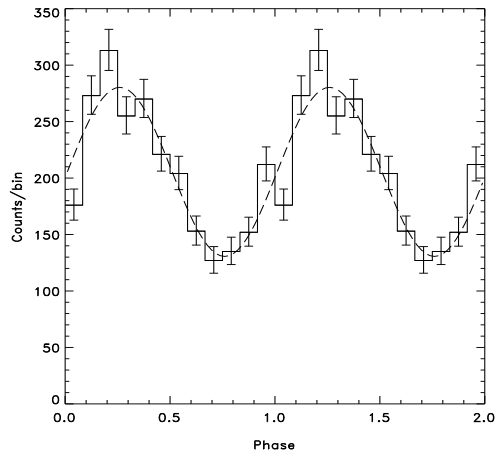}
    \caption{\label{mini_outburst_two_b} A 12 bin pulse profile of \axpeinstein\ for events with PHA values in the range 4--50 
             ($\sim$ 2-20 keV) from any PCU-2 detector layer gathered during a 52 s time period starting at the onset of first 
              burst of the Jan. 11, 2010 mini-outburst. Two cycles are shown for clarity. A sine fit (dashed line) is superposed to guide the eye and to demonstrate 
              that the pulsed emission is highly symmetric. A $13\sigma$ signal is detected during this very short time window indicating 
              the presence of strong pulsating tail emission.}
  \end{center}
\end{figure}

\subsubsection{Mini-outbursts}
\label{sect:mini_outbursts}

A detailed look at the decaying tail of the Jan. 22, 2009 outburst shows a large deviation from the global decay
trend near MJD 54920 (Mar. 30, 2009). Initially, we believed that we were dealing with a processing error/anomaly, but
a deeper investigation revealed a period of enhanced emission of about 450 s with superposed at least three short duration 
bursts (see Fig.\ref{mini_outburst_one_rate}). This behaviour was seen in both PCU-1 and PCU-2, which were both operational 
during the observation in question, thus excluding an instrumental (non-celestial) cause of the observed phenomenon.
The period of enhanced emission was preceded by a short duration burst.

We searched for more mini-outbursts and found one on Jan. 11, 2010 (MJD 55207.8845 TT, see Fig. \ref{mini_outburst_two_a}). 
Enhanced emission is seen for a period of about 130 s with superposed two bursts with a separation of $\sim 11$ s between the burst maxima.

An interesting feature of this enhanced emission period is that the burst trail is highly pulsed. We found a pulsed signal strength of
$\sim 13\sigma$ (see Fig. \ref{mini_outburst_two_b}) for the burst-cleaned emission confined in a time window of only 52 s starting at the onset of the first burst for the 4--50 PHA range ($\sim 2-20$ keV) using all detector layers of PCU-2, the only detector operational during the observation. This proves that the enhanced emission is originating from \axpeinstein.
Next, we revisited the enhanced emission period observed earlier by the PCA on MJD 54920, but in this case we did {\it not} find any evidence 
for highly increased pulsed emission.

Similar behaviour has been detected by \citet{kaneko10} using Fermi GBM data during a period of enhanced persistent emission of 
$\sim 150$ s duration starting only 70 s after the first GBM trigger at the onset of the Jan. 22, 2009 outburst\footnote{Note, that the 
Fermi GBM (all-sky monitor) detected the onset of the Jan. 22, 2009 outburst about 39 minutes earlier than Swift BAT \citep[][only $\frac{1}{6}$ of the sky is visible by the BAT at any instant]{gronwall09}}.
\begin{table*}[t]
\caption{Incoherent timing models for \axpeinstein\ as derived from RXTE PCA monitoring data.}
\label{ineph_table}
\begin{center}
\begin{tabular}{lcccllll}
\hline
Entry &  Start &  End  &   t$_0$, Epoch   & \multicolumn{1}{c}{$\nu$}   & \multicolumn{1}{c}{$\dot\nu$}               & 
\multicolumn{1}{c}{$\ddot\nu$}                  & \multicolumn{1}{c}{$\dddot{\nu}$}\\
 \#   &  [MJD] & [MJD] &     [MJD,TDB]    & \multicolumn{1}{c}{[Hz]}    & \multicolumn{1}{c}{$\times 10^{-12}$ Hz/s}  & \multicolumn{1}{c}{$\times 10^{-18}$ Hz/s$^2$}  &  \multicolumn{1}{c}{$\times 10^{-25}$ Hz/s$^3$}             \\
\hline\hline
\\
1         & 54743 & 54845 & 54780.0     & 0.482728(2)    & -22.3(11)               & -2.5(3)                    & 15(4)   \\
2         & 54896 & 55121 & 54995.0     & 0.482496(2)    & -11.0(3)                & -0.44(8)                   & 0       \\
3         & 55120 & 55404 & 55270.0     & 0.482201(2)    & -18.1(5)                & -0.50(9)                   & 1.2(3)  \\
\vspace{-2mm}\\
\hline\hline
\end{tabular}
\end{center}
\end{table*}
Also, \citet{mereghetti09} reported on pulsating tails lasting several seconds from two bright bursts occuring on Jan. 22, 2009 detected above 80 keV by the SPI ACS aboard the INTEGRAL satellite.

It is worth mentioning that given the sparse RXTE sampling (about 5 ks of observation time per week, i.e. 0.8\% of the time, is devoted to 
observe \axpeinstein) we very likely miss a lot of such events i.e. periods of enhanced persistent emission possibly preceded by a burst event. If such events are accompanied by (micro) glitches, the (very) noisy timing behaviour (see also Sect. \ref{coherent_models}), inhibiting the construction of phase coherent timing models over time stretches longer than about 30 days, can easily be explained.

\subsection{Evolution of the pulse frequency: incoherent measurements}
\label{evolution_pulse_frequencies}

In each RXTE sub-observation of \axpeinstein\ a coherent pulsed signal at a rate of $\sim 0.482$ Hz \citep[see][]{camilo07} 
could be detected in the barycentered time series. We used a $Z_1^2$-test \citep{buccheri1983} search in a small, typically 5
independent Fourier steps\footnote{a Fourier step is defined by $\Delta\nu_{\hbox{\scriptsize IFS}}=1/\tau$, in which $\tau$ represents the time span of the data period} wide, window around the predicted pulse frequency.
The restricted search yielded for each sub-observation a best estimate of the rotation rate at the gravity point
of the sub-observation. Because the $Z_1^2$-test is distributed as a $\chi^2$ for $2\times 1$ degrees of freedom a $1\sigma$ error 
estimate on this optimum value can easily be derived by determining the intersection points of the measured $Z_1^2$-test 
distribution near the optimum with the value of $Z_{\hbox{\scriptsize 1,max}}^2-2.296$.

The pulse frequency measurements for \axpeinstein\ from Oct. 3, 2008 till December 25, 2010 are shown as bold data points along with their $1\sigma$ errors in the middle panel of Fig. \ref{evolution_flux_nu_pflux}. The measurements are relative to a coherent model 
(see later in Sect. \ref{coherent_models}), which could be constructed for the time period MJD 54855-54884 (entry \#3 in Table \ref{eph_table}), just after the onset of the Jan. 22, 2009 outburst, using both RXTE PCA and Swift XRT data.

\subsubsection{Incoherent models}
\label{incoherent_models}

It is clear from the behaviour of the pulse frequency displayed in Fig. \ref{evolution_flux_nu_pflux} (middle panel) that we can
identify three time periods in which the spin behaviour can be described with only a limited number of parameters. Apparent discontinuities
in the timing behaviour (glitches) are present near MJD 54853 i.e. the onset of the Jan. 22, 2009 outburst, and near MJD 55121 (Oct. 17, 2009).

The frequency behaviour in each of the three time intervals was determined by fitting the measured pulse frequencies and their $1\sigma$ uncertainties by a simple timing model (Taylor series) with either 3 or 4 timing parameters ($\nu,\dot\nu,\ddot\nu[,\dddot\nu]$): \begin{equation}\nu_{inc}(t)=\nu + \dot\nu \cdot (t-t_0) + \frac{1}{2}\ddot\nu\cdot (t-t_0)^2 +\frac{1}{6}\dddot\nu\cdot (t-t_0)^3 \label{eq:incoh}\end{equation}
The three incoherent timing solutions $\nu_{inc}(t)$ are listed in Table \ref{ineph_table} and shown in Fig. \ref{evolution_flux_nu_pflux} (middle panel) as solid lines together with the $1\sigma$ uncertainty band (dashed lines), all relative to the coherent timing model \#3
(see Table \ref{eph_table}).
\begin{table*}[t]
\caption{Phase-coherent ephemerides for \axpeinstein\ as derived from RXTE PCA and Swift XRT data.}
\label{eph_table}
\begin{center}

\begin{tabular}{lccclllcc}
\hline
Entry &  Start &  End  &   t$_0$, Epoch   & \multicolumn{1}{c}{$\nu$}   & \multicolumn{1}{c}{$\dot\nu$}               & \multicolumn{1}{c}{$\ddot\nu$}                  & $\Phi_{0}^{a}$  & Validity range\\
 \#   &  [MJD] & [MJD] &     [MJD,TDB]    & \multicolumn{1}{c}{[Hz]}    & \multicolumn{1}{c}{$\times 10^{-12}$ Hz/s}  & \multicolumn{1}{c}{$\times 10^{-18}$ Hz/s$^2$}  &                &  \multicolumn{1}{c}{(days)}   \\
\hline\hline
\\
1         & 54743 & 54779 & 54743.0     & 0.48277818(3)    & -6.48(5)               & -6.46(4)                   & 0.4613   & 37\\
2         & 54819 & 54845 & 54819.0     & 0.48264979(4)    &-18.67(4)               &  0.0 (fixed)               & 0.7382   & 27\\
\vspace{-2mm}\\
3         & 54855 & 54884 & 54856.0     & 0.48259525(1)    & -5.12(2)               &  0.0 (fixed)               & 0.0710   & 30\\
4         & 54855 & 54890 & 54856.0     & 0.48259518(3)    & -4.91(5)               & -0.23(4)                   & 0.0711   & 36\\
\vspace{-2mm}\\
5         & 54977 & 55012 & 54977.0     & 0.48251286(6)    &-11.05(9)               & -0.18(7)                   & 0.5144   & 36\\
6         & 55213 & 55243 & 55229.0     & 0.48226226(3)    &-15.91(2)               & -1.07(12)                  & 0.1590   & 31\\
\vspace{-2mm}\\
\hline\hline
\multicolumn{9}{l}{$^a$ $\Phi_{0}$ is the phase offset to be applied to obtain consistent phase-alignment (see Eq. \ref{eq:phase} in Sect. \ref{coherent_models}) }\\
\end{tabular}
\end{center}
\end{table*}

\subsubsection{Coherent models: phase-connected\\ ephemerides}
\label{coherent_models}
\begin{figure}[ht]
  \begin{center}
    \includegraphics[height=7.0cm,width=7.5cm,angle=0,bb=20 160 555 665,clip=]{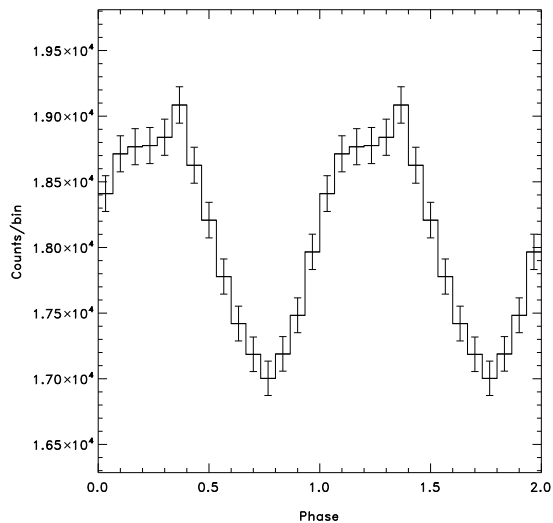}
    \caption{\label{correlation_template} The RXTE PCA pulse profile (15 bins) for the PHA range 4-10 (measured energies $\lesssim 4.2$ keV) 
              used as template in the Time-of-Arrival (TOA) correlation analysis.}
  \end{center}
  \vspace{-0.75cm}
\end{figure}

Very accurate timing models (ephemerides) taking into account every rotation of the neutron star can be obtained by employing phase-coherent 
timing techniques. This requires a set of pulse arrival times (ToA's) which results from a correlation analysis of an instantaneous profile with
a template correlation profile. For the phase-coherent timing techniques used in this work we refer to Section 4.1 of \citet{kuiper09}.
In this study the correlation template is based on RXTE PCA events with measured energies $\lesssim 4.2$ keV, corresponding to the 4-10 PHA
range, because in this range the morphology of the pulse profile is least (almost not) affected by transient morphology changes (see Sect.\ref{xray_pp_morph}), 
which are prominently present in other bands at higher energies $\gtrsim$ 4.2 keV.
It also facilitates the inclusion of ToA's based on Swift XRT observations, because then the PCA and XRT effective energy bands are fully overlapping. The PCA correlation template (PHA 4-10; $\lesssim 4.2$ keV) is shown in Fig. \ref{correlation_template} adopting 15 bins, and it is
based on RXTE observations performed a couple of days after the Jan. 22, 2009 outburst.

Reliable phase-coherent timing models are given in Table \ref{eph_table}. During a 37-days period after the onset of the Oct. 2008 outburst
\axpeinstein\ was intensively monitored by both RXTE and Swift (from MJD 54742-54764, only). For this period entry \#1 of Table \ref{eph_table}
specifies the model parameters based on the combined PCA/XRT ToA's set. These are consistent with those published by \citet{ng2011} and \citet{israel2010}. Note, the large negative value for $\ddot\nu$, indicating a rapidly increasing spin-down rate.

Entries \#3 and \#4 of Table \ref{eph_table} show the timing parameters for a period of intensive Swift and RXTE monitoring (just) after the second outburst in Jan. 2009, and are valid for a 30-days (2 parameters) and 36-days (3 parameters) time period, respectively. The timing models are consistent with those reported in \citet{ng2011} and \citet{bernardini2011}. Comparing these models with the one (entry \#2) valid for the period just before the Jan. 2009 outburst i.e. MJD 54819-54845\footnote{Two Swift XRT observations in WT mode have been performed on Jan. 4 and 13, 2009, respectively. The pulse profile resulting from the combination of these XRT observations adopting the timing 
model (entry \#2 of Table \ref{eph_table}) based on PCA data only, shows the expected alignment.}, indicates that a strong glitch, mainly 
in $\dot\nu$, occurred at a time consistent with the onset of the Jan. 2009 outburst. More details on this apparent glitch are shown later in Sect.
\ref{glitchjan09}. Finally, the validity of coherent timing models \#5 and \#6 of Table \ref{eph_table}, based on PCA data only, have been 
verified using independent Swift XRT data taken in WT-mode. Folding these Swift data yielded consistent pulse alignment. 

Also, for other time intervals with interval lengths of typically 30-35 days we could generate phase-coherent models from PCA data, mostly 
using 2-d ($\nu,\dot{\nu}$) optimization schemes applied for a $2\sigma$ range around predicted ($\nu,\dot{\nu}$) values estimated from an
incoherent timing model (see Table \ref{ineph_table}).
However, often due to the lack of independent contemporaneous data from other X-ray instruments e.g. Swift XRT\footnote{Most of the time the Swift 
XRT operated in PC mode, inadequate for timing studies of \axpeinstein.} and given the sparse sampling of 
the RXTE observations, we can not securely identify the found ``best'' model as the true underlying rotation model.
These models, not listed in Table \ref{eph_table}, are still useful for combining RXTE observations in spectral studies covering several sub-time 
intervals.

Finally, for this source phase coherence is typically lost after a $\sim 30-35$ day time period due to the sparse sampling of both RXTE and Swift 
observations and the high level of timing noise \citep[see e.g. ][]{camilo08}.


\subsection{Glitches}
\begin{figure}[t]
  \begin{center}
    \includegraphics[height=6.5cm,width=7.5cm,angle=0,bb=50 185 510 640,clip=]{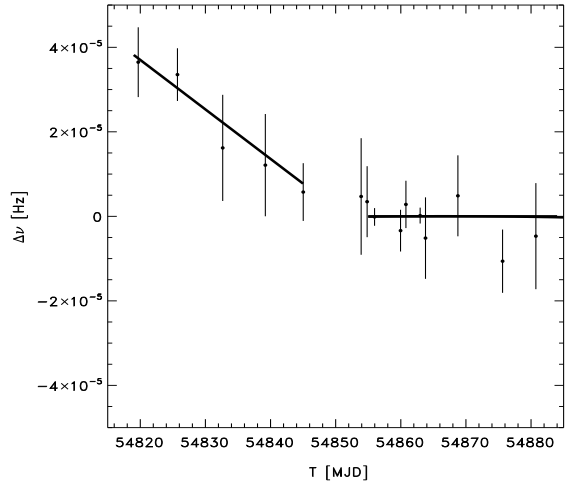}
    \caption{\label{glitch_nu_behvr}A zoom-in of the rotation frequency behaviour around the Jan. 2009 outburst. Frequency measurements are given by
             the data points, while the solid lines refer to coherent timing models \#2 and \#3 of Table \ref{eph_table}. All frequency values are relative to predictions by model \#3. An apparent glitch occurred somewhere between MJD 54845.02 and 54853.91, very likely simultaneously with the onset of the Jan. 2009 outburst at MJD 54853.035 (TDB).}
  \end{center}
  \vspace{-0.00cm}
\end{figure}
\subsubsection{The characteristics of the glitch associated with the Jan. 22, 2009 outburst}
\label{glitchjan09}
In order to study the Jan. 2009 glitch in detail we adopted standard glitch fitting techniques as employed
e.g. in the pulsar timing software package {\tt tempo2} vrs 1.9\footnote{See http://www.atnf.csiro.au/research/pulsar/tempo2}. We use the following, commonly used, expression for the evolution of the frequency crossing a glitch:
\begin{equation}
\nu(t) = \nu_0(t) + \Delta \nu_p   
                  + \Delta \nu_d e^{-(t-t_g)/\tau_d}
                  + \Delta \dot{\nu} (t-t_g),
\end{equation}
where $\nu_0(t)$ is the frequency of the pulsar prior to the glitch occuring at time $t_g$, $\Delta \nu_p$ is the frequency jump, 
that is permanent, while $\Delta \nu_d$ is the part that (exponently) decays at time scale $\tau_d$, and finally $\Delta \dot{\nu}$ 
specifies the jump in the frequency time derivative. The initial (at time $t=t_g$) frequency jump $\Delta \nu$ is given by 
$\Delta \nu_p+\Delta \nu_d$.

The frequency behaviour prior and after the glitch (see Fig.\ref{glitch_nu_behvr}) does not show any recovery term, and therefore we abandon the decaying part of the frequency jump. Assuming, that the glitch occurred at the time of the onset of the Jan. 2009 outburst i.e. MJD 54853.035 
(TDB) we derived the following values for the relative frequency jump $\Delta\nu/\nu$ of $(1.9\pm1.6)\cdot 10^{-6}$, and frequency derivative 
jump $\Delta\dot{\nu}/\dot{\nu}$ of $-0.69\pm0.07$. Note, that the value for the frequency jump is not significant at face value, in contrast 
to the frequency derivative jump value of $\Delta{\dot\nu}=+(1.30\pm0.14)\cdot 10^{-11}$ Hz/s, indicating a dramatic decrease of the spin-down
rate. Such an instantaneous decrease in spin-down rate, from $(-18.8\pm 0.6)\cdot 10^{-12}$ Hz/s to $(-5.12 \pm 0.02)\cdot 10^{-12}$ Hz/s, has not been seen before from any magnetar (see the discussion section for a comparison with earlier reports on magnetar timing, Sect.\, \ref{sect_disc}).

\begin{figure}[h]
  \begin{center}
    \includegraphics[height=5.5cm,width=7.5cm,angle=0,bb=35 255 555 575,clip=]{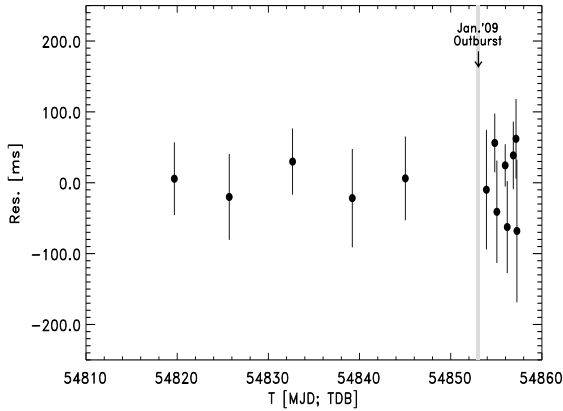}
    \caption{\label{glitch_model} Arrival time residuals (in ms) across the range MJD 54819-54857 for a fit based on a timing model crossing 
             the glitch epoch, that was hold fixed at the onset of the Jan. 2009 outburst at MJD 54853.035 (TDB).}
  \end{center}
  \vspace{-0.00cm}
\end{figure}

The arrival time residuals for the best fitting timing model crossing the glitch, thus including the frequency and frequency derivative jumps,
are shown for the range MJD 54819-54857 in Fig.\ref{glitch_model}, and indicate a proper description of the measured ToA's by the 
glitch-crossing timing model.
\subsubsection{October 2009 glitch}
\label{glitchoct09}
From the middle panel of Fig.\ref{evolution_flux_nu_pflux}, showing the frequency measurements with superposed the three incoherent timing models, 
another discontinuity in the timing behaviour of \axpeinstein\ can be discerned near MJD 55121 (Oct. 17, 2009). Unfortunately, the RXTE (and Swift) observation sampling is very sparse, such that a detailed study is impossible. From the incoherent models (plus uncertainties; see Table \ref{ineph_table}) near MJD 55121 we can determine that also for this glitch a strong $\Delta\dot{\nu}/\dot{\nu}$ jump occurred of size ($-0.89\pm 0.19$), similar to the Jan. 22, 2009 glitch.
More surprisingly, in this case the glitch is not accompanied by a radiative outburst (cf. top and middle panels of Fig.\ref{evolution_flux_nu_pflux}).
\begin{table*}[ht]
\caption{Definitions of the RXTE observation time segments.\label{rxte_segments}}
\setlength{\tabcolsep}{4pt}
{\footnotesize
\begin{center}

\begin{tabular}{ccccc}
\hline
 \textbf{Segment}     & \textbf{Begin/End Date} & \textbf{Start/End} & \textbf{Duration}   & \textbf{PCU-2 / 0--4} \\
 \textbf{nr.}         & \textbf{(yyyy/mm/dd)}   & \textbf{(MJD)}     & \textbf{(days)}     & \textbf{exposure (ks)}     \\
\hline
 1                    & 2008-10-03 / 2008-10-29  & 54743  - 54768    & 26 & \,64.007 / 116.874 \\
 2                    & 2008-11-01 / 2009-01-14  & 54771  - 54845    & 75 & \,68.783 / 120.092 \\
\\[-5pt]
 3                    & 2009-01-22 / 2009-02-01  & 54853  - 54863    & 11 & \,84.094 / 130.199\\
 4                    & 2009-02-06 / 2009-05-03  & 54868  - 54954    & 87 & \,79.662 / 149.528\\
 5                    & 2009-05-11 / 2009-06-22  & 54962  - 55004    & 43 & \,40.614 / \,73.717\\
 6                    & 2009-07-14 / 2009-10-08  & 55026  - 55112    & 87 & \,45.331 / \,77.139\\
 7                    & 2009-10-16 / 2009-12-26  & 55120  - 55191    & 72 & \,47.714 / \,81.113\\
 8                    & 2010-01-17 / 2010-03-09  & 55213  - 55264    & 52 & \,34.472 / \,62.683\\
 9                    & 2010-03-17 / 2010-05-12  & 55272  - 55328    & 57 & \,38.368 / \,71.177\\
 10                   & 2010-05-17 / 2010-07-27  & 55333  - 55404    & 72 & \,45.784 / \,83.488\\
 11                   & 2010-08-04 / 2010-12-25  & 55412  - 55555    &144 & 109.704 / 193.350\\
\hline\hline
\\[-5pt]
\multicolumn{5}{l}{PCU-2 and PCU 0--4 exposures are screened adopting default selection criteria,} \\
\multicolumn{5}{l}{removing short duration bursts and taking into account detector break-downs.} \\
\end{tabular}
\end{center}}

\end{table*}
\subsection{Pulsed flux evolution at X-rays below $\sim$ 10 keV}
\label{pulsed_flux}

Using either the verified (see Table \ref{eph_table}) or non-verified (see Sect.\ref{coherent_models}) coherent timing models
we phase-folded barycentered PCA event arrival times according to \begin{equation}\Phi(t)=\nu\cdot (t-t_0) + \frac{1}{2}\dot\nu\cdot (t-t_0)^2+\frac{1}{6}\ddot\nu\cdot (t-t_0)^3 - \Phi_0 \label{eq:phase},\end{equation} to obtain pulse-phase
distributions. ($\nu,\dot\nu,\ddot\nu$) represent the pulse frequency, first time derivative of the frequency and second time derivative of the frequency, respectively, while $t_0$ is the epoch of the ephemeris.
Consistent X-ray phase alignment is obtained by subtracting $\Phi_0$ as shown in Eq. \ref{eq:phase}.

Next, for every PCA sub-observation we fitted the pulse-phase distribution, based on events from PCU-2 with PHA values in the range [4,27] (measured energy $\la 10$ keV; all Xenon detector layers used), with a model consisting of a constant plus a truncated Fourier series (the fundamental and 2 harmonics).
This yielded for every PCA sub-observation the number of pulsed excess counts which in turn was converted to a (pulsed) count rate using the known exposure time. These PCU-2 (pulsed) count rates for the [4,27] PHA range are shown in the bottom panel of Fig. \ref{evolution_flux_nu_pflux} for the time window MJD 54743-55555 (Oct. 3, 2008 -- December 25, 2010).

If we compare the total and pulsed X-ray flux measurements (both $\sim$2-10 keV) by RXTE PCA (cf. top and bottom panels of Fig.\ref{evolution_flux_nu_pflux}), then a completely different evolution is shown, indicating that the pulsed component is (at most) weakly related to the unpulsed or DC component. Most strikingly, the radiative outburst of the pulsed flux in October 2008 is stronger than that in January 2009, while the total flux outburst in January 2009 is much more intense than that in October 2008.

Concentrating on the pulsed-flux evolution we notice that, i) in October 2008 the pulsed flux maximum is reached with a delay of about 10 days with respect to the onset of the Oct. 3, 2008 outburst, which is consistent with the findings by \citet{ng2011} and \citet{scholz2011}, and ii) after the start of the Jan. 22, 2009 outburst the pulsed flux levels appeared to vary by a factor $\sim$2 for about two weeks, then more or less stabilize, and a period of gradual decrease sets in lasting from MJD 54868 up to MJD 55555.
Assuming a linear decay over the full period, we find a good reduced $\chi_{\nu}^2$ of $66.11/79 = 0.84$ with a best fit decay rate of $(-6.15 \pm 0.66)\times 10^{-4}$ c/(s day), estimated for the $\sim 2-10$ keV band of PCU-2, which means a drop of about $54\%$ over the full 688 day period (see dashed line in the bottom panel of Fig.\ref{evolution_flux_nu_pflux}).


\section{Pulse profile morphology evolution at X-rays below $\sim$ 30 keV}
\label{xray_pp_morph}
\begin{figure*}[h]
\centerline{\includegraphics[height=18cm,width=12cm,angle=0]{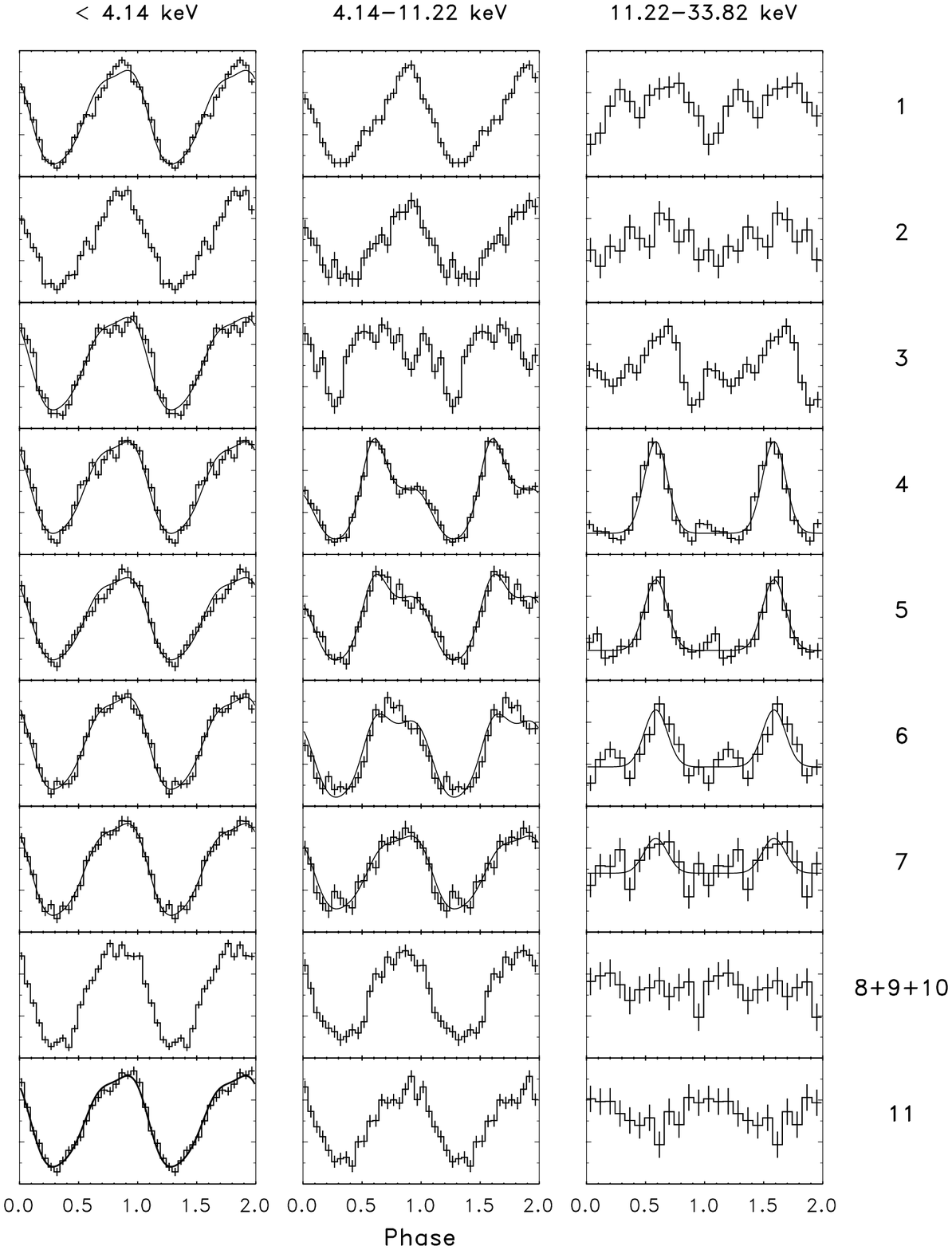}}
  \caption{\label{morph_vs_e_t} The evolution of the morphology of the (PCA) X-ray pulse profiles in 3 different energy bands (see headings at the top of the figure). Time-segment identifiers are shown on the right-hand side, and are defined in Table \ref{rxte_segments}. In particular, segments 1 and 2 refer to the time period between the Oct. 3, 2008 and Jan. 22, 2009 outbursts. Segment-3 represents a time period of only 11 days duration starting just after the Jan. 22, 2009 outburst, during which rapid and drastic morphology and flux changes occur.}
\end{figure*}
In order to study in detail the evolution of the pulse profile morphology in various X-ray bands, we combined the data from several RXTE/PCA sub-observations into 11 time segments spread over the time period from Oct. 3, 2008 up to Dec. 25, 2010 (MJD 54743-55555).
Details of these RXTE observation time segments can be found in Table \ref{rxte_segments}. The first two time segments refer to RXTE observations performed between the Oct. 3, 2008 and Jan. 22, 2009 outbursts, while the other segments have start times beyond the Jan. 22, 2009 outburst (see middle panel of Fig.\,\ref{evolution_flux_nu_pflux}).

After folding each barycentered PCA event time stamp according to Eq. \ref{eq:phase} from burst and detector break-down free periods we generated for each of the time segments event matrices with $(60 \times 256)$ elements by binning the pulse-phase range [0,1] into 60 phase bins for all 256 PHA channels. 
In this process we used data from {\em all} three xenon layers of each PCU, which considerably improves the signal-to-noise ratio for energies above $\sim 10$ keV. This allows us to better characterize the hard X-ray ($>10$ keV) properties. In Fig.\,\ref{morph_vs_e_t} the
pulse profiles of \axpeinstein\ are shown for all 11 time segments adopting three different X-ray bands: $< 4.14$ keV (soft band, PHA 4-10), $4.14-11.22$ keV (medium band, PHA 11-27) and $11.22-33.82$ keV (hard band, PHA 28-80). Drastic morphology changes as a function of time are shown for the hard and medium energy bands, especially for the transitions from segment 2 to 3, and 3 to 4. Apparently, the dramatic changes of physical conditions after the Jan. 22, 2009 event (the timing glitch and radiative outburst occur just prior to the start of segment-3) are responsible for these phenomena 

Noteworthy is that the hard X-ray bands of segment 1 and 2 already show pulsed emissions at the $4.7\sigma$ and $2.9\sigma$ significance levels (applying $Z_2^2$-tests), respectively. Therefore, the period just after the Oct. 3, 2008 outburst represents the first time period in which from this source pulsed emission has been detected significantly at energies above $\sim 10$ keV. 

From Fig. \ref{morph_vs_e_t} it is also clear that the soft X-ray pulse profiles ($< 4.14$ keV) exhibit least variability and remain relatively stable. This stability was the main reason for choosing the soft X-ray band for the ToA correlation analysis (see Sect. \ref{coherent_models}). 

\subsection{Evolution of a transient hard X-ray pulse above $\sim$ 10 keV}
\label{xray_pp_hard}
\begin{figure}[t]
  \begin{center}
    \includegraphics[height=7.5cm,width=7.5cm,angle=0,bb=65 180 525 630,clip=]{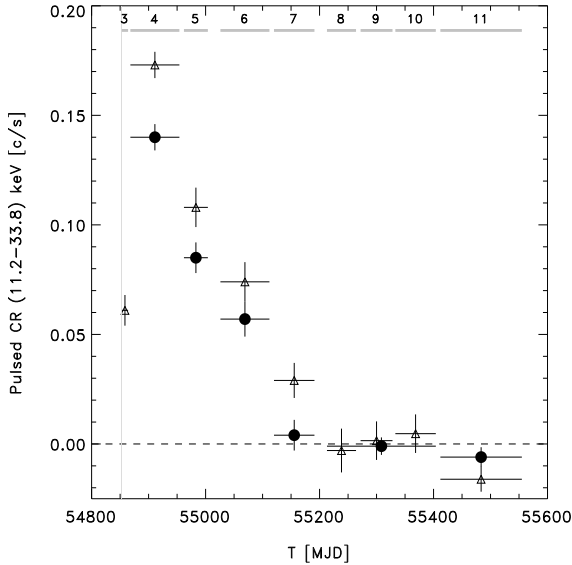}
    \caption{\label{hard_pflux_evol} Count-rate evolution since the Jan. 22, 2009 outburst (shown as vertical grey line) of the transient hard X-ray pulse (triangles) for energies $\sim 11.2-33.8$ keV (see Sect. \ref{xray_pp_hard}). The selected PCA time segments are indicated near the top of the figure.
    The hard pulsed component reaches its maximum in segment 4, which starts 15 days after the onset of the Jan. 22, 2009 outburst and lasts 87 days. Next, this component fades to undetectable levels within $\sim 350$ days from the onset. The fading trend of this component is also clearly visible at lower energies ($\sim 4.1-11.2$ keV; dots). All errors are statistical only ($1\sigma$). 
}
  \end{center}
\end{figure}
In the hard X-ray band of segment-4 (see Fig. \ref{morph_vs_e_t} right panel labeled 4) a completely new, relatively sharp, emission feature pops up near phase 0.6 after a short transition phase during segment-3, which lasts only 11 days. This feature gradually fades and is not detectable anymore beyond segment-7.
To obtain the best statistics for this new component in the hard X-ray band we stacked all data from segment 4 up to and including 7, and fitted the resulting pulse phase distribution (in 60 bins) with a model composed of a Gaussian (3 free parameters) and a (flat) background (1 free parameter). The reduced $\chi_{\nu}^2$ of the best fit is $68.21/(60-4) = 1.22$ indicating that the model provides an adequate description of the data. The resulting centroid and width (FWHM=$2.354\sigma$) of the best fit Gaussian are $0.587(4)$ and $0.23(1)$, respectively.

In order to follow the evolution of the fading hard X-ray pulse we fitted the hard X-ray pulse profiles of segments 3 to 11 in terms of this ``best fit'' Gaussian (fixed position and width) with free scale and a constant. From the fits we derived the number of excess counts associated with the Gaussian component, which together with the (screened) exposure time for the combination of PCU 0--4 (see 8$^{\hbox{\scriptsize{th}}}$ column of Table \ref{rxte_segments}) can be translated to pulsed count rates. 

The quality of the nine fits was good/acceptable except for segments 3 and 4, yielding for both reduced $\chi_{\nu}^2$ values of $\sim 1.8$ for $60-2=58$ degrees of freedom. For segment-4, a slight shift to the left of the Gaussian template provides an excellent fit and a consistent count rate. 
In the case of segment-3 the observed profile is deviating too much from a Gaussian shape. A fit with a truncated Fourier series (fundamental plus 2 harmonics) to determine the unpulsed level yielded a good fit (reduced $\chi_{\nu}^2=59.54/(60-7)=1.12$).
Now, the pulsed excess counts derived in a phase window [0.426-0.748], centered on $0.587$ containing a 90\% fraction of the ``best fit'' Gaussian, above this differently derived unpulsed level resulted in a pulsed count rate $\sim 1.6$ times higher than that estimated by the Gaussian extraction method. 
Segment-3 seems to represent a transition period starting after the glitch epoch till a configuration/geometry is reached with the relatively narrow hard X-ray pulse at phase $\sim0.59$.
In Fig.\,\ref{hard_pflux_evol} the hard X-ray pulsed count rate, adopting the Gaussian extraction method, is shown versus time since the Jan. 22, 2009 outburst. It is clear, that the hard X-ray pulsed component fades to undetectable levels within about 350 days from the onset of the Jan. 22, 2009 outburst.

Also, in spite of the difficulties in reliably estimating the hard X-ray pulsed count rate for segment-3, we can securely state that the maximum emission of the hard X-ray pulsed component is delayed with respect to the Jan. 22, 2009 outburst. This is supported by non-uniformity significance estimations of the pulse phase distributions in the hard X-ray band using $Z_2^2$ statistics for segment 3 and 4. Although both segments have comparable exposure times (see Table \ref{rxte_segments}), we detect for segment-3 only a $9.2\sigma$ pulsed signal, while for segment-4 we find a $27.5\sigma$ signal (cf. Fig.\,\ref{morph_vs_e_t} right panels 3,4).

\subsection{Modelling the pulse profiles below 10 keV}
\label{xray_pp_medium}
The evolution of the pulse shape in the middle panel of Fig.\,\ref{morph_vs_e_t} (4.14-11.22 keV band) suggests that the profiles of segments 4 and beyond are composed of two components: 1) a relatively stable one with a shape compatible with that at lower energies (see left panel of Fig.\,\ref{morph_vs_e_t}) and 2) a transient Gaussian-like feature centered near phase 0.6, the phase of the transient pulse above 10 keV.

In order to model the evolution of the profile shape in the medium band from segment 4 and beyond we generated now also a template for the soft X-ray band ($<4.14$ keV) using data from the time periods covering segments 8--11, a year after the Jan. 2009 timing glitch. This soft X-ray template has been obtained applying a truncated Fourier series fit, using the fundamental and two harmonics, of the combined $<4.14$ keV pulse phase distributions of segment 8/9/10 and 11 (see lower two left panels
of Fig.\,\ref{morph_vs_e_t}). The resulting template is superposed on the low-energy pulse phase distribution of segment 11 (see bottom left corner of Fig. \ref{morph_vs_e_t}).

Next, we fitted all $<4.14$ keV pulse phase distributions in terms of a constant and the soft X-ray template with free normalization, and obtained good/reasonable (even for segment-3) fits for all segments except 1 and 2 (most of the fits are superposed on the distributions shown in the left panel of Fig.\,\ref{morph_vs_e_t}). Therefore, we conclude that the pre- Jan. 22, 2009 outburst soft X-ray profile differs significantly from that of the post-Jan. 22, 2009. The post-outburst model clearly underestimates the pulsed emission near the maximum, and overestimates the contribution near phase 0.6 (see left top panel of Fig.\,\ref{morph_vs_e_t}).

Finally, the medium-energy profiles from segments 4 and beyond were fitted with a model composed of a constant, the soft X-ray and
hard X-ray (Gaussian) templates both with free scale. We obtained acceptable fits, i.e. $\chi_{\nu}^2 \la 1.36$ for 57 degrees of freedom, for all time segments except for segment-6 ($\chi_{\nu}^2 \simeq 1.82$).
From the scale factors of the best fits and exposure times the count rates for both the soft X-ray and hard X-ray components can be determined. For the hard X-ray (Gaussian) component these rates are added in Fig.\,\ref{hard_pflux_evol}, and show a similar fading trend as visible at higher X-ray energies, meaning that the hard Gaussian like transient pulse extends towards lower X-ray energies.
For segment-4 $38\pm 2 \%$ of the total pulsed emission in the medium-energy band can be attributed to the hard X-ray Gaussian component, while for segment-5 and 6 these numbers are $24 \pm 2\%$ and $15 \pm 2\%$, respectively.
 

\section{Broad-band spectral evolution: total and pulsed X-ray emissions up to $\sim$ 300 keV}
\label{sect_gen_spc_evol}

For our spectral analyses of the total and pulsed emissions, we addressed the broad energy range from $\sim$ 1 keV up to $\sim$ 300 keV, exploiting the different capabilities of instruments aboard three high-energy observatories: RXTE, Swift and INTEGRAL. Data from the non-imaging RXTE instruments PCA and HEXTE have been
used to obtain spectral information on the pulsed emission over a broad energy range, $\sim 2.5 -250$ keV, while data from the
Swift XRT instrument in PC-mode have been analysed to derive the total emission spectrum at soft X-rays (0.2-10 keV).
Using INTEGRAL ISGRI data spectral information has been obtained for both the total and pulsed emission of \axpeinstein\ over the
20-300 keV energy band. Finally, during the early phase of the decay of the Jan. 2009 outburst JEM-X1 (3-30 keV) detected \axpeinstein\
up $\sim 15$ keV, and through imaging analysis spectral information of the {\em total} emission could even be obtained for energies between 10 and 15 keV, a range not/hardly accessible for other high-energy instruments with imaging capabilities.

All our spectral results have been corrected for interstellar absorption adopting a Hydrogen column density N$_{\hbox{\scriptsize H}}$  of $3.12\times 10^{22}$ cm$^{-2}$ \citep[see e.g.][ slightly above the total Galactic H\,{\scriptsize I} column density estimated to be in the range $(1.81-2.25) \times 10^{22}$ cm$^{-2}$]{gelfand07,halpern08}, and thus correspond to unabsorbed source spectra. Note, that these corrections are only effective for photon energies below $\sim 5$ keV, and thus it is not necessary to apply these to HEXTE and ISGRI data.
\subsection{RXTE PCA and HEXTE pulsed X-ray spectra}
\label{X_total-pulsed_spectra}
\begin{figure}[t]
  \begin{center}
    \includegraphics[height=6cm,width=7.5cm,angle=0,bb=100 265 490 550,clip=]{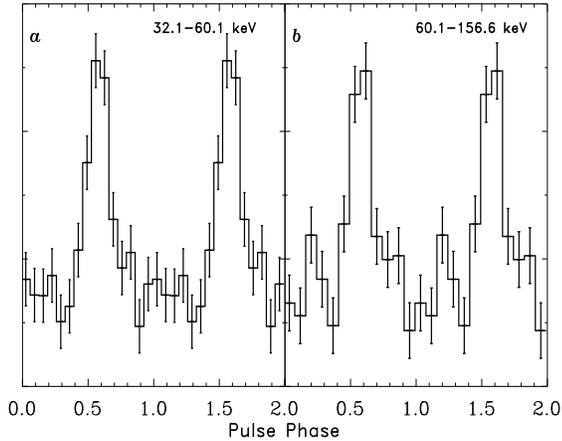}
    \caption{\label{hexte_prof}HEXTE pulse-profiles for the 32.1-60.1 ($10.6\sigma$; panel a) and 60.1-156.6 ($8.6\sigma$; panel b) keV bands combining
                               data from RXTE time segments 4---7. The Gaussian shaped hard X-ray profile extends up to 
                               $\sim 150$ keV.
    }
  \end{center}
  \vspace{-0.3cm}
\end{figure}
\begin{figure}[t]
  \begin{center}
    \includegraphics[height=7.5cm,width=7.5cm,angle=0,bb=50 150 545 660,clip=]{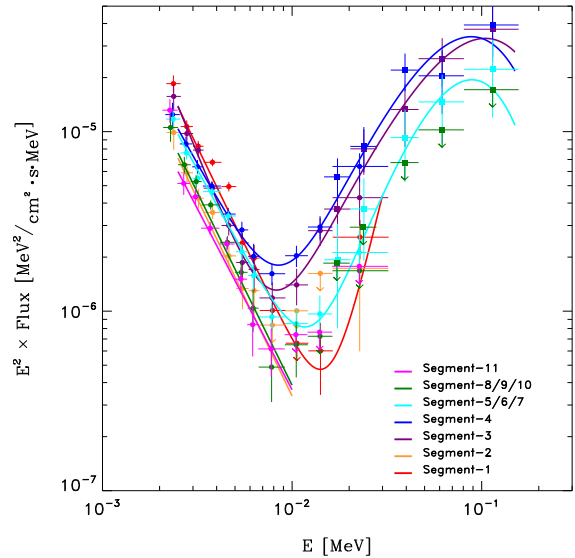}
    \caption{\label{pulspc_evol}Spectral evolution from Oct. 2008 to Dec. 2010 of the pulsed (unabsorbed) emission of \axpeinstein\ in the
                                2.5-150 keV band as measured by RXTE PCA (filled circles) and HEXTE (filled squares) for time segments 1 -- 11 (see
                                Table \ref{rxte_segments} and middle panel Fig.\,\ref{evolution_flux_nu_pflux}).
                                Power-law plus cutoff power-law model fits are superposed for segments 3, 4 and 5/6/7 (see Table \ref{pulsed_flux_tab}), during which significant pulsed emission was detected by HEXTE above $\sim 15$ keV, otherwise only power-law-model fits are shown below 10 keV. The pulsed hard X-ray emission ($\ga 10$ keV) becomes maximal during segment-4, showing therefore a delayed increase with respect to the onset of the Jan. 2009 outburst, before a gradual decline sets in.
    }
  \end{center}
  \vspace{-0.3cm}
\end{figure}

Analogous to the procedure followed for the PCA to obtain event matrices (see Sect.\ref{xray_pp_morph}) we generated for the HEXTE data for all 11 time segments two dimensional (burst-free) event distributions sorting on pulse-phase and energy. First, we verified that for the combined RXTE time segments 4--7 we also detect with HEXTE at hard X-rays the same Gaussian-shaped high-energy pulse as measured with the PCA below $\sim$ 30 keV. Fig. \ref{hexte_prof} shows the HEXTE pulse profiles for the energy bands 32.1--60.1 keV and 60.1--156.6 keV in which this Gaussian profile is found at the same pulse phase, with non-uniformity significances 10.6 and $8.9\sigma$, respectively.

Next, we derived pulsed excess counts in user selected energy (PHA) bands for both PCA and HEXTE data by fitting
a truncated Fourier series (using the fundamental plus 2 harmonics) to the pulse-phase distributions measured in these bands. The fit function minimum plus associated error are used to determine the number of pulsed excess counts i.e. the number of counts above this minimum (=unpulsed, i.e. DC plus background level) for each user-selected energy band.

For the PCA we constructed time-averaged energy response matrices for each PCU separately taking into account the different (screened) exposure times of the involved PCU's during the time period of interest. For this purpose we used the {\it ftools version 6.4} programs {\it pcarsp} and {\it addrmf}. To convert PHA channels to measured energy values, $E_{\hbox{\scriptsize PHA}}$, for PCU combined/stacked products we also generated a weighted PCU-combined energy response matrix.

For HEXTE we employed cluster A and B energy-response matrices separately, taking into account the different screened on-source exposure times and the reduction in efficiency in case of off-axis observations. The on-source exposure times for both clusters have been corrected for the considerable dead-time effects.

To translate the derived pulsed excess counts to photon flux values we employed these response matrices in (forward folding) spectral fitting procedures adopting simple underlying photon emission models like a power-law, a combination of two power-laws or a power-law plus an (exponential) cutoff power-law.
\begin{table*}[t]
\caption{Characteristics of the spectral fits, adopting a combination of a power-law and a cutoff power-law, to RXTE PCA and HEXTE (2.5-150 keV) {\em pulsed\/} flux data.\label{pulsed_flux_tab}}
\setlength{\tabcolsep}{3pt}
{\footnotesize
\begin{center}
\begin{tabular}{ccccccccc}
\hline
\\[-5pt]
\textbf{RXTE}      & $E_0$   &  $k_1$          & $\Gamma_1$ &  $k_2$         & $\Gamma_2$ & $E_c$   &  $F_E^{2-10}$    &  $F_E^{20-150}$\\
\textbf{segm.}     & (keV)   &  $\times 10^6$  &            &  $\times 10^5$ &            & (keV)   &                  &                \\
\hline
\\[-5pt]
3                  & 12.9295 & $1.10\pm0.11$ & $-4.63^{+0.08}_{-0.07}$ & $1.49\pm0.32$ & $+0.37^{+0.20}_{-0.28}$ & $44.0\pm 10.4$ & $1.47^{+0.06}_{-0.07}$ & $6.9^{+1.2}_{-1.0}$          \\
\\[-5pt]
4                  & 12.8760 & $2.68\pm0.21$ & $-3.92^{+0.06}_{-0.06}$ & $1.86\pm0.33$ & $+0.59^{+0.19}_{-0.25}$ & $33.8\pm 5.9$  & $1.28^{+0.04}_{-0.05}$ & $7.5^{+0.9}_{-1.0}$          \\
\\[-5pt]
5/6/7              & 12.8875 & $2.50\pm0.15$ & $-3.92^{+0.05}_{-0.05}$ & $0.43\pm0.15$ & $+1.55^{+0.26}_{-0.42}$ & $25.3\pm 4.8$  & $1.18^{+0.03}_{-0.04}$ & $4.1^{+0.9}_{-0.9}$          \\
\\[-5pt]
\hline
\\[-6pt]
\multicolumn{9}{l}{1) The (photon flux) fit model is: $F_{\gamma}(E_{\gamma})= k_1 \cdot (E_{\gamma}/E_0)^{\Gamma_1} + k_2 \cdot (E_{\gamma}/E_0)^{\Gamma_2} \cdot \exp(-E_{\gamma}/E_c)$  } \\
\\[-6pt]
\multicolumn{9}{l}{2) The normalisations $k_1$ and $k_2$ are in units ph/cm$^2$s keV at pivot energy $E_0$.  } \\
\\[-6pt]
\multicolumn{9}{l}{3) The unabsorbed energy fluxes $F_E^{2-10}$ and $F_E^{20-150}$ are in units $10^{-11}$ erg/cm$^2$s} \\
\\[-6pt]
\multicolumn{9}{l}{4) Quoted errors are for a 68.27\% confidence level ($1\sigma$)}\\
\end{tabular}
\end{center}}
\end{table*}

The results of the spectral fits to the combined PCA/HEXTE data sets\footnote{We applied the following energy-independent normalization factors of 0.912 (PCA) and 1.087 (HEXTE) to the spectral measurements in order to obtain consistent high-energy Crab pulsed fluxes \citep[see Sect. 3.4 of ][]{kuiper06}.} for the pulsed emission of all 11 segments along with the (reconstructed) flux measurements are shown in a $\nu F_{\nu}$ representation in Fig. \ref{pulspc_evol} for the range $\sim 2.5-150$ keV.
In this figure the HEXTE results for segments 1, 2 and 11 are left out for clearity, because the non-detections in the different energy bands correspond to $2\sigma$ upper-limit levels comparable to those shown for segment-8/9/10. However, HEXTE did detect pulsed emission
($3.6\sigma$) in its integral energy band, $\sim 15-250$ keV, for segment-1, confirming the PCA findings (see Sect. \ref{xray_pp_morph}).

Below $\sim 10$ keV the pulsed spectra are all, irrespective of the time segment, very soft with photon indices $\Gamma$ in the range $-$[3.9--4.2] (except for segment-3, $\Gamma \simeq -4.6$). Also, a significant decrease by a factor of $\sim 2$ in normalization is shown for the time period between the two outbursts, moving from segment-1 to 2. Then, promptly after the Jan. 2009 outburst a strong increase in normalization by a factor of $\sim 2$ is observed, moving from segment-2 to 3, followed by a gradual decrease down to the level shown for segment-11. The whole picture for energies below $\sim 10$ keV is consistent with the behaviour in the coarse ($\sim 2-10$ keV) energy band shown in the bottom panel of Fig.\ref{evolution_flux_nu_pflux}.

Above $\sim 10$ keV the variations are more drastic. Since the initial activation of pulsed high-energy emission after the Oct. 2008 outburst this component reached undetectable levels during segment-2, but increased dramatically after the Jan. 2009 outburst, reaching
eventually a maximum during segment-4, after which a gradual decrease sets in resulting to undetectable levels beyond segment-7.

During the periods with significant pulsed hard X-ray emission the best spectral description over the full 2.5-150 keV band consists of 
a combination of a power-law plus a cutoff power-law model (see Table \ref{pulsed_flux_tab} for quantitative information). The improvement relative to a model composed of two power-laws is about $3.5\sigma$, considering the combination of the three periods, segments 3, 4 and 5/6/7, during which significant pulsed hard X-ray
emission has been detected.

For the cutoff power-law model component, $F(E_{\gamma})= k \cdot (E_{\gamma}/E_0)^\Gamma \cdot \exp(-E_{\gamma}/E_c)$ with $k$ the normalization in
ph/cm$^2$ s keV, $E_0$ the pivot energy, $\Gamma$ the power-law (photon) index and $E_c$ the cutoff energy, the photon index at a specific energy $\Gamma^* := \frac{d\ln F}{d\ln E_{\gamma}}$ is given by $\Gamma^*=\Gamma - E_{\gamma}/E_c$. Therefore, for this model component the maximum in a $\nu F_{\nu}$ spectral representation (maximum power per energy decade) is found at $E_{\gamma}^{\max} = (\Gamma + 2)\cdot E_c$. For the spectral fits of segments 3, 4 and 5/6/7 the best fit parameters for $\Gamma$ and $E_c$ yield $\nu F_{\nu}$ maxima for energies in the range 90--105 keV with an uncertainty of about 20 keV. We stress that our data donot constrain the extrapolation of the spectra, the spectral shape, above these energies.

\axpeinstein\ exhibits in time segment 4, extending over 87 days, the maximal pulsed flux at energies above 10 keV. An important parameter for theoretical modelling is the time between the onset of the outburst/timing glitch and the epoch at which
the maximal flux of the non-thermal transient component is reached. We used the high count rate in the RXTE PCA to divide segment 4 in three time intervals to study the spectral evolution in more detail. The normalisation of the spectra appeared maximal in two
intervals between 37 and 101 days after the onset of the Jan. 2009 outburst. We refer to this delay as 70$\pm$30 days. 

\subsection{Swift XRT soft X-ray spectra of the total emission}
\label{swift_xrt_tot}
To study the evolution of the {\em total} (=pulsed plus unpulsed) emission spectrum of \axpeinstein\ at soft X-rays (0.5-10 keV) since the Jan. 2009 outburst we used Swift XRT data obtained in PC-mode, allowing full 2D-imaging information (see Sect.\,\ref{instr_swift}). This information is necessary to get rid of the emission from the dust scatter rings \citep[][]{tiengo10} which were prominently present during the early part of the decay phase after the Jan. 2009 outburst.

The Swift observations selected for this analysis are chosen such, that these cover (sometimes approximately) time periods for which hard X-ray INTEGRAL observations have been performed (see Table \ref{obsint_table} and Sect.\,\ref{instr_swift} for the INTEGRAL and Swift observations, respectively),
supplemented with some recent observations (e.g. those with obs. identifiers 00090404019, Sept. 28, 2010, and 00090404027, Feb. 25, 2011) to follow the late time evolution.

We selected events from a $60\arcsec$ circular aperture centered on \axpeinstein\ using (default) cleaned event lists. For this selection we executed procedure {\tt xrtmkarf} (vrs. 0.5.8 embedded in the {\tt HEASOFT} instrument specific software analysis environment) to obtain appropriate sensitive area information, stored in so-called ancillary response files (arf), which takes into account the reduction in sensitive area for off-axis observations and the finite angular size of the source extraction region (about 90\% of the source counts falls within a $60\arcsec$ circle centered on the source). The energy redistribution matrix (rmf) used was {\tt swxpc0to12s6\_20070901v011.rmf}. 

We also selected a background region composed of an annular region with a
$180\arcsec$ inner- and $240\arcsec$ outer radius centered on \axpeinstein, 7 times larger than the source region.
From this we estimated the number of background counts in our source region, which was $\la 1\%$ of the total number of counts confined in our source region, and therefore no background correction has been made in subsequent spectral fitting procedures.

Sorting the selected (source region) events on energy we fitted the resulting event distribution in measured energy space to a certain model photon
spectrum attenuated by interstellar absorption (see Sect.\,\ref{sect_gen_spc_evol}) in a forward folding procedure using the appropriate response
information ({\tt arf} and {\tt rmf} response files). As model (source) spectrum a combination of a black-body (BB) and power-law (Pl) was used which can adequately reconstruct (empirically) the measured spectral distributions below 10 keV. These model fits were also succesfully applied and discussed for \axpeinstein\ by e.g. \citet[]{gelfand07,ng2011,bernardini2011,scholz2011}.
\begin{table*}[t]
\caption{Characteristics of the spectral fits, adopting a combination of a power-law and a black-body, to solely Swift XRT-PC (0.5-10 keV) {\em total} flux data. Note, however, that the $\Gamma$ values change, when the INTEGRAL flux values above 20 keV are included in the fits (see Table \ref{totspec_prop}). \label{xrt_bb_pl}}
\setlength{\tabcolsep}{3pt}
{\footnotesize
\begin{center}
\begin{tabular}{cccccccc}
\hline
\\[-5pt]
\textbf{Swift XRT} &  Date       & MJD range           &  $kT$             & $\Gamma$         & $F_E^{1-10}$    & $F_E^{1-10}$   & color\\
\textbf{obs.}      &             &                     &  (keV)            &                  &  (Pl-model)     &  (BB-model)    & in Fig.\ref{totspc_evol}\\
\hline
\\[-6pt]
00030956034        & 29/01/2009  & 54860.005---54860.343 &  $0.606\pm 0.007$ & $-1.14\pm 0.04$  &  $5.07\pm 0.50$ & $4.48\pm 0.32$ & orange\\
\\[-6pt]
00030956039        & \,\,\,4/02/2009  & 54866.702---54866.974 &  $0.564\pm 0.006$ & $-1.32\pm 0.04$  &  $4.41\pm 0.37$ & $4.26\pm 0.31$ & purple\\
\\[-6pt]
00030956051        & 19/08/2009  & 55062.069---55062.150 &  $0.775\pm 0.018$ & $-1.99\pm 0.16$  &  $1.71\pm 0.45$ & $2.52\pm 0.36$ & dark blue\\
\\[-6pt]
00090404003        & 12/04/2010  & 55298.509---55298.938 &  $0.728\pm 0.014$ & $-2.84\pm 0.20$  &  $0.93\pm 0.22$ & $1.36\pm 0.16$ & aqua\\
\\[-6pt]
\hline
\\[-6pt]
\multicolumn{8}{l}{1) The (photon flux) fit model is: $F(E_{\gamma})= \alpha_{bb}\cdot E_{\gamma}^2 / (\exp(E_{\gamma}/kT)-1) + \alpha_{pl} \cdot E_{\gamma}^{\Gamma}$} \\
\\[-6pt]
\multicolumn{8}{l}{2) The energy fluxes $F_E^{1-10}$ are unabsorbed (N$_{\hbox{\scriptsize H}}=3.12\cdot 10^{22}$ cm$^{-2}$) and in units $10^{-11}$ erg/cm$^2$s for the 1--10 keV band} \\
\\[-6pt]
\multicolumn{8}{l}{3) Quoted errors are for a 68.27\% confidence level ($1\sigma$)}\\
\\[-6pt]
\end{tabular}
\end{center}}
\end{table*}

The spectral fit parameters adopting this composite BB+Pl model are given in Table \ref{xrt_bb_pl} and the flux measurements for the 0.5--10 keV range reconstructed from spectral fits are shown in Fig.\,\ref{totspc_evol}. Energy fluxes provided in Table \ref{xrt_bb_pl} are unabsorbed and specified for the 1--10 keV band to facilitate comparisons with the results of \citet{scholz2011}. 
All fits have acceptable reduced $\chi^2$-values near 1 assuming a fixed N$_{\hbox{\scriptsize H}}$ of $3.12 \times 10^{22}$ cm$^{-2}$. The fit values for the BB temperature kT and power-law index $\Gamma$ as a function of time since the start of the outburst are consistent with those reported by e.g. \citet{bernardini2011} and \citet{scholz2011}. Table \ref{xrt_bb_pl} also confirms a clear softening of the power-law index with time and a decreasing flux below 10 keV (also visible in the spectral shapes below 10 keV in Fig.\,\ref{totspc_evol}). This hardness-intensity correlation has been discussed previously, e.g. by \citet{scholz2011}, however, our spectral fits to the combined Swift-INTEGRAL spectra over the broad energy band 1--300 keV provide a different, alternative decomposition and hardness-intensity correlation (Sect. \ref{integralswifttotsp}).
\begin{figure*}[t]
  \begin{center}
    \includegraphics[height=16cm,width=8cm,angle=90]{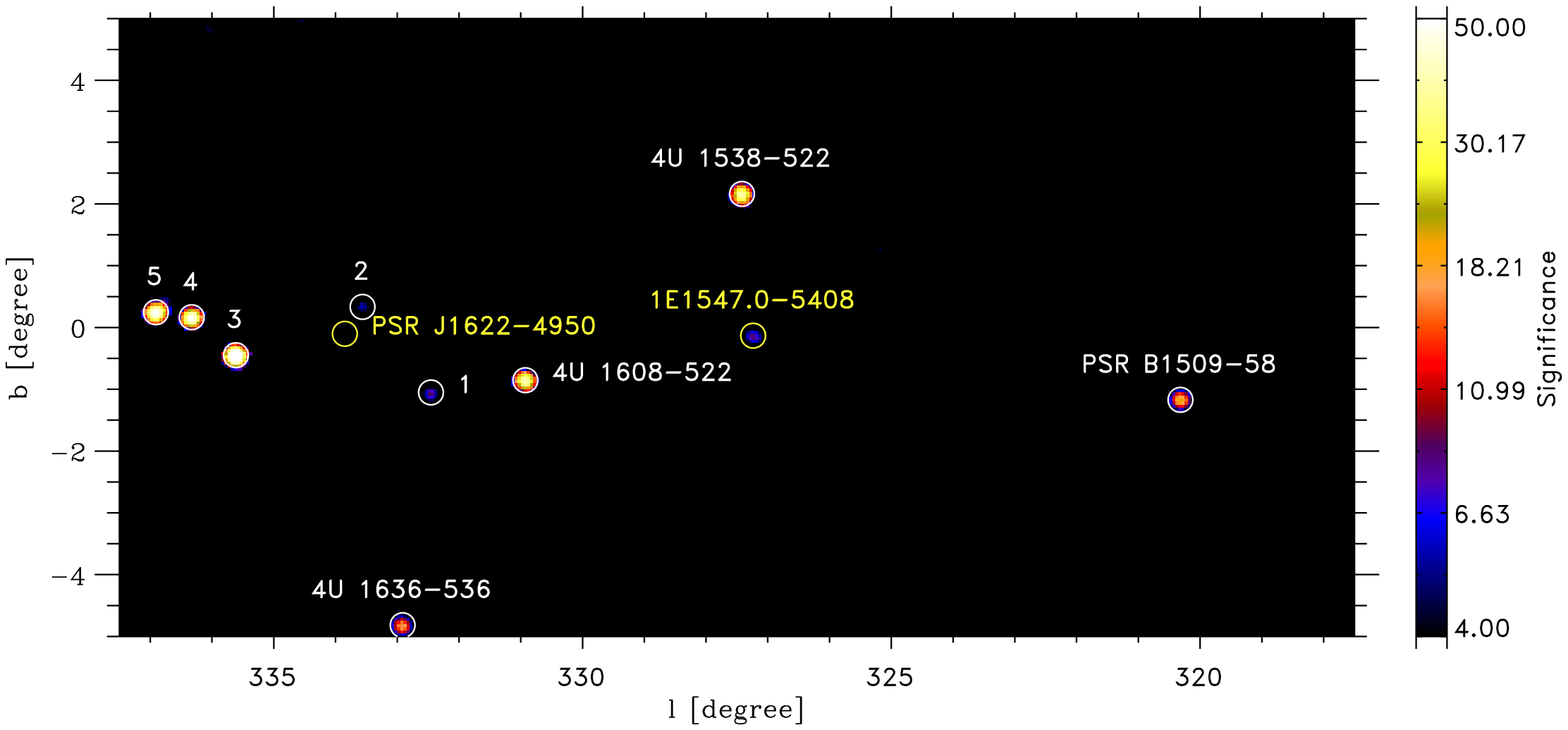}
    \caption{\label{isgri_map} INTEGRAL ISGRI significance map for the 20-150 keV band of a $20^{\circ} \times 10^{\circ}$ field centered  on \axpeinstein\ based on observations performed during INTEGRAL revolutions 899--912 (Feb. 23 -- Apr. 3, 2010), more than a year after the Jan. 2009 outburst. The effective exposure on \axpeinstein\ is $\sim 717.7$ ks, and its detection significance is still $9.5\sigma$. The location of another radio-loud magnetar, PSR J1622-4950, not detected by INTEGRAL, is also indicated by a yellow circle. Other high-energy sources detected by ISGRI during revolutions 899--912 are indicated either by name or number. Those labelled with a number are: [1] IGR J16207-5129, [2] AX J1619.4-4945, [3] IGR J16318-4848, [4] AX J1631.9-4752 and [5] 4U 1636-536.
    }
  \end{center}
  \vspace{-0.3cm}
\end{figure*}
\subsection{INTEGRAL soft $\gamma$-ray total emission spectra}
\label{integraltotsp}
The ISGRI observations (science windows) selected for our study all have observation dates between Oct. 3, 2008 and Apr. 3, 2010 and instrument pointings within $14\fdg5$ from the \axpeinstein\ sky location. This ensures that (a part of) the detector plane is illuminated by the target. The resulting list is further screened on erratic count rate variations, indicative for particle-induced effects due to Earth-radiation-belt passages or solar-flare activities, by inspecting visually the count rate in 20-30 keV band versus time. Science windows showing erratic count-rate variations are excluded for further analysis. Moreover, we ignored time intervals during which bursts (from any source in the field of view) occurred. This selection is particularly important for INTEGRAL revolutions 767--772 during which copious numbers of bursts from \axpeinstein\ have been detected. The set of INTEGRAL revolutions used along with the effective on-axis exposure times is given in Table \ref{obsint_table}.
\begin{figure}[t]
  \begin{center}
    \includegraphics[height=7.5cm,width=7.5cm,angle=0,bb=45 150 545 660,clip=]{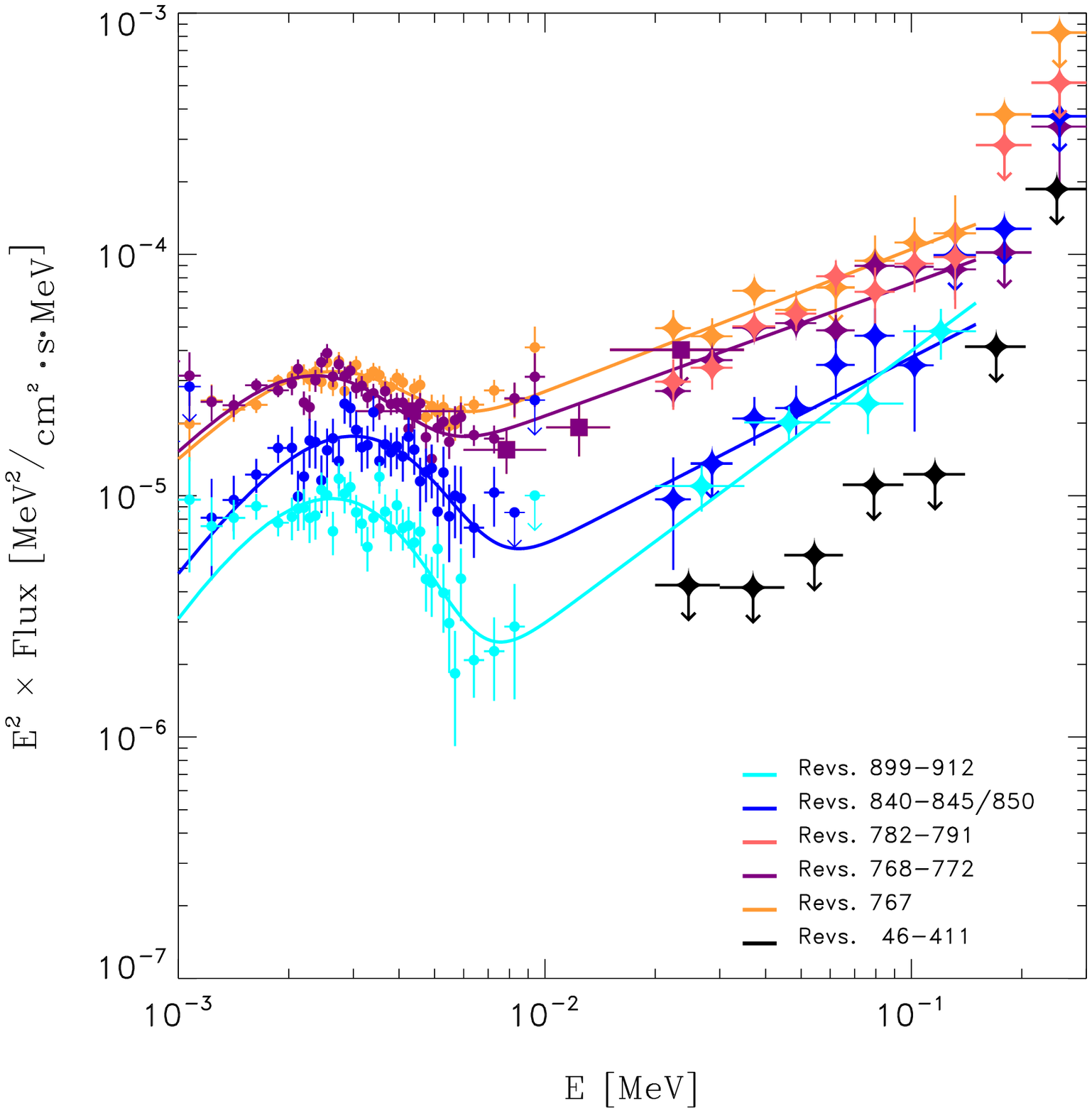}
    \caption{\label{totspc_evol}Spectral evolution after the timing glitch/outburst in Jan. 2009 up to April 2010 of the total (unabsorbed) emission of 
                                \axpeinstein\ in the 1-300 keV band as measured by Swift-XRT (PC-mode; filled circles), INTEGRAL JEM-X (filled
                                 squares) and ISGRI (filled diamonds). Black-body plus power-law-model fits for the combined 
                                 fluxes are shown. The different colors indicate the INTEGRAL observations (revolutions) with (nearly) 
                                 contemporaneous Swift observations. Note, that for the period covering Revs. 782--791 no XRT observations have 
                                 been performed in PC-mode. Black 2$\sigma$ upper-limits are shown for the combined exposure of $\sim 4$ Ms of all 
                                 INTEGRAL observations (prior to the Oct. 2008 outburst) from March 2003 to Feb. 2006.
    }
  \end{center}
  \vspace{-0.15cm}
\end{figure}

The {\em total} emission spectrum at soft gamma rays / hard X-rays above 20 keV can be derived by exploiting the arcminute imaging capability of ISGRI. We generated skymaps for each selected science window (see Sect.\,\ref{instr_integral}) in different energy bands covering the 20-300 keV energy window. From these maps sky mosaics were assembled using dedicated imaging software tools embedded in the OSA 7.0 analysis environment. We chose in most cases a logarithmically binned energy grid with 10 bins covering the 20-300 keV range, but also 4 logarithmic energy bins over the 20-150 keV range when we expected low flux levels for \axpeinstein\ (e.g. for INTEGRAL Revs. 899--912).
The final product of the imaging analysis provides the count rate, its variance, exposure and significance for all energy bands over the (deconvolved) mosaiced sky field. As an example Fig.\,\ref{isgri_map} shows the significance map (20-150 keV) in a wide field centered on \axpeinstein\ combining observations from INTEGRAL revolutions 899-912, performed more than a year after the Jan. 2009 outburst. \axpeinstein\ is clearly detected in this mosaic at a level of $\sim9.5\sigma$. 

\begin{table*}[t]
\caption{\label{totspec_prop} Spectral properties of the {\em total} emission of \axpeinstein\ across the 1-300 keV range for various INTEGRAL observation periods (orbital revolution numbers, see Table\,\ref{obsint_table}) after the timing glitch in Jan. 2009, derived by fitting a black-body plus 
power-law model to (nearly) contemporaneous INTEGRAL ISGRI and Swift XRT data. INTEGRAL revolutions 46-411 represent a total of 4 Ms of observations of \axpeinstein\ before March 2006.}
{\footnotesize
\begin{center}
\begin{tabular}{cccccc}
\hline
\textbf{INTEGRAL}   & \textbf{Flux}          & $\Gamma$       &\textbf{kT}        &\textbf{BB-Flux}$^{[1]}$      &\textbf{Pl-Flux}$^{[1]}$    \\
\textbf{Obs. period}    & \textbf{20-150 keV}    &                &\textbf{(keV)}     &\textbf{2-10 keV}     &\textbf{2-10 keV}   \\
\hline\hline
767             & $(2.52\pm0.37)$E-10    & $-1.41\pm0.06$  & $0.621 \pm 0.008$ & $(2.51\pm 0.22)$E-11 & $(4.49\pm 0.23)$E-11\\
768-772         & $(1.85\pm0.18)$E-10    & $-1.45\pm0.04$  & $0.566 \pm 0.007$ & $(2.35\pm 0.19)$E-11 & $(3.66\pm 0.17)$E-11\\
\,\,\,\,\,\,782-791$^{[3]}$ & $(2.15\pm0.14)$E-10    & $-1.27\pm0.11$  & ...               & ...                  & ...                 \\
840-850         & $(0.84\pm0.27)$E-10    & $-1.22\pm0.10$  & $0.739 \pm 0.014$ & $(2.23\pm 0.27)$E-11 & $(0.91\pm 0.17)$E-11\\
899-912         & $(0.80\pm0.22)$E-10    & $-0.87\pm0.07$  & $0.654 \pm 0.010$ & $(1.18\pm 0.11)$E-11 & $(0.35\pm 0.06)$E-11\\
\,\,\,\,\,\,\,\,\,\,\,46-411$^{[2,3]}$          & $<0.15$E-10    & $-1.35$ (fixed) & ...               & ...                  & ...\\
\hline
\hline
\multicolumn{6}{l}{$^{[1]}$All fluxes are unabsorbed adopting N$_{\hbox{\scriptsize H}}=3.12\times 10^{22}$ cm$^{-2}$ and in units erg/cm$^2$s}\\
\multicolumn{6}{l}{$^{[2]}$Upper limit at $2\sigma$ confidence}\\
\multicolumn{6}{l}{$^{[3]}$Only ISGRI data used adopting power-law model}\\
\end{tabular}
\end{center}
}
\end{table*}

The total emission spectrum of \axpeinstein\ can be derived from the count-rate and variance maps by extracting the (dead-time corrected) rates and uncertainties at the location of \axpeinstein.
These values are normalized to the count rates measured for the total (nebula and pulsar) emission from the Crab in similar energy bands, using Crab calibration observations (in $5\times5$ dither mode) performed during INTEGRAL revolution 102. From the ratios and the photon spectrum of the total emission from the Crab, we can derive the total high-energy photon spectrum of \axpeinstein\ (pulsed and any unpulsed point source component). For the total Crab photon spectrum we used the broken-power-law spectrum derived by \citet{jourdain08} based on INTEGRAL-SPI observations of the Crab at energies between 23 and 1000 keV. The latest Crab cross calibrations between SPI and IBIS-ISGRI provided consistent results.

Fig.\,\ref{totspc_evol} shows the ISGRI flux measurements (20-300 keV) for different sets of observations performed between Jan. 24, 2009 and April 3, 2010 (Revs.\,767--912; see Table \ref{obsint_table}).
For the (only) 100 ks ToO observation performed during 8--10 Oct., 2008 (Rev-731; a couple of days after the Oct. 3, 2008 outburst) no significant flux values can be derived with $2\sigma$ upper-limits (not shown in Fig.\,\ref{totspc_evol}) consistent with the pulsed flux detections by RXTE PCA and HEXTE.
We added in Fig.\,\ref{totspc_evol} also the $2\sigma$ upper-limits on the \axpeinstein\ flux derived from a deep 4 Ms mosaic \citep[see e.g.][]{kuiper08} using INTEGRAL observations performed between revolution 46 and 411 (March 2, 2003 -- Feb. 24, 2006) targeting on PSR J1617-5055, which is located within $5\fdg5$ from \axpeinstein. These upper-limits are at least 10 times lower than the flux levels reached during INTEGRAL revolutions 767--791 in which the hard spectral tail is detected most significantly.

During the intensive monitoring campaign (Revs. 767--772) of \axpeinstein\ the source was also for $\sim264.8$ ks within $3\fdg75$ of the JEM-X1 pointing axis. In the combination of these observations JEM-X1 clearly detected \axpeinstein\ at a $8.6\sigma$ level between 3.04 and 15.04 keV. Flux values are obtained using similar procedures as applied for ISGRI i.e. count rates are expressed in Crab (total) flux units and subsequently converted to flux units, however, now we have to take into account the different column densities along the line of sight for the Crab (N$_{\hbox{\scriptsize H}}$ of $3.2\times 10^{21}$ cm$^{-2}$) and \axpeinstein\ (N$_{\hbox{\scriptsize H}}$ of $3.12\times 10^{22}$ cm$^{-2}$).
The four JEM-X1 flux measurements (3 detections and 1 upper-limit) are shown in Fig.\,\ref{totspc_evol} as filled squares.

\subsection{INTEGRAL and Swift XRT total-emission spectral evolution}

\label{integralswifttotsp}
To extend the spectral coverage down to $\sim1$ keV we combined the ISGRI(/JEM-X1) observations with the (nearly) contemporaneous Swift XRT observations performed in PC-mode (see Sect.\,\ref{swift_xrt_tot}). This allowed us to study the spectral shape and spectral evolution over the broad 1--300 keV interval. We found that a fit with a black-body plus (single) power-law model to the flux measurements over the total energy range provided a statistically good fit. No fudge factors were used in these fits to absorb calibration uncertainties between different instruments. The model fits are shown in Fig.\,\ref{totspc_evol}. The spectra of the two late-time Swift-XRT observations of Sept. 28, 2010 and Feb. 25, 2011 are  not shown for clarity, because these are fully overlapping with the spectrum (aqua-coloured) of the observation performed on April 12, 2010 slightly after INTEGRAL revolutions 899--912.

Characteristics of these XRT-ISGRI combined spectral fits are listed in Table\,\ref{totspec_prop}. The first thing to notice is that also for this broad energy band two spectral components, a BB plus a power-law model, appear to be sufficient to fit the flux values \citep[see also][]{bernardini2011}. The fit values for the BB temperature $kT$ (Table\,\ref{totspec_prop}) are close to those derived for solely 0.5--10 keV band Swift-XRT data (Table \ref{xrt_bb_pl}), however, the power-law photon indices $\Gamma$ in the third column of Table\,\ref{totspec_prop} are very different, ranging from $\sim -1.4$ after the Jan. 2009 glitch/outburst to $\sim-0.9$ almost 14 months later, while the 20-150 keV flux is fading by a factor of $\sim$ 3. In agreement with this, \citet{denhartog2009}, using $\sim 200$ ks of INTEGRAL (near real time) ToO data taken 2--8 days after the Jan. 2009 glitch/outburst, and \citet{enoto2010a}, using Suzaku data taken $\sim 7$ days after the Jan. 2009 glitch/outburst, consistently reported a power-law index of $\sim -1.5$ during the first half of INTEGRAL observation period Revs. 767-772 (see Table \ref{obs_table}). 
Furthermore, \citet[][Suzaku data]{enoto2010a} and \citet[][Chandra and INTEGRAL data]{bernardini2011} also derived for energies $\sim$ 0.5--100 keV within statistical uncertainties the same fit result as given in Table\,\ref{totspec_prop} for epochs in INTEGRAL Revs. 767-772. Thus, the power-law component resulting from the fits below 10 keV, is not required when the broad energy range is considered. The ``new'' hard power-law component explains the total hard X-ray emission above $\sim 10$ keV and also contributes significantly to the emission below 10 keV. Moreover, the index is hardening with decreasing flux, the opposite trend to what is visible in Table \ref{xrt_bb_pl} and discussed in e.g. \citet{scholz2011} for spectral fits restricted to energies below 10 keV.

\begin{figure}[t]
  \begin{center}
    \includegraphics[height=7.0cm,width=7.5cm,angle=0,bb=85 180 565 630,clip=]{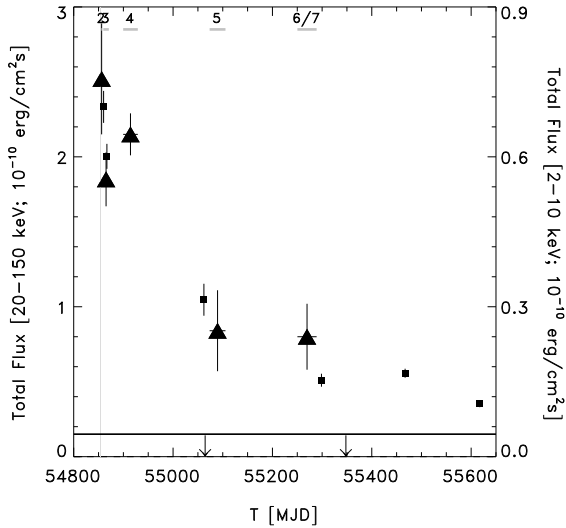}
    \caption{\label{totflux_evol} Evolution of the total (unabsorbed) energy flux of \axpeinstein\ since the Jan. 2009 outburst in two different energy bands: 2-10 keV (Swift-XRT; solid squares; right flux scale) and 20-150 keV (ISGRI; filled triangles; left flux scale). The onset of the Jan. 2009 outburst is indicated by a vertical line, and the INTEGRAL observation segments (see Table\,\ref{obsint_table}) are shown in the top part of the figure. The horizontal line represents the $2\sigma$ upper limit for the 20-150 keV flux as derived from INTEGRAL observations performed during Revs. 46-411 (before March 2006).
    }
  \end{center}
  \vspace{-0.15cm}
\end{figure}

The fading trends of the total fluxes at soft X-rays and hard X-rays since the Jan. 2009 outburst are shown in Fig.\,\ref{totflux_evol}.
The total (unabsorbed) soft X-ray flux is the sum of the black-body and power-law components (last 2 columns of Table\,\ref{totspec_prop}). The power-law contribution at soft X-rays is negligible after April 12, 2010. Namely, late-time Swift XRT measurements of the total soft X-ray spectrum of \axpeinstein\ demonstrate that a single black-body component is sufficient to describe the spectrum adequately.
The (late-time) flux measurements of Swift observations 0009040419 (Sept.\,28, 2010) and 00090404027 (Feb.\,25, 2011) are added in Fig.\,\ref{totflux_evol}. The corresponding (unabsorbed) 2-10 keV fluxes and temperatures are ($1.67(9)\cdot 10^{-11}$ erg/cm$^2$s, kT=0.671(5)) and ($1.06(7)\cdot 10^{-11}$ erg/cm$^2$s, kT=0.600(6)), respectively.

It is evident from Figs.\,\ref{totspc_evol} and \ref{totflux_evol} that the total source fluxes are maximal during the first INTEGRAL ToO observation after the Jan. 2009 glitch/outburst epoch (Rev. 767, starting 2 days after the Jan. 2009 outburst) for energies below 10 keV as well as for the soft gamma-rays/hard X-rays (20--150 keV). It is remarkable that for the latter {\em total\/} non-thermal emission we do not see evidence that maximal luminosity is reached about 70 $\pm$ 30 days after the glitch (in RXTE and INTEGRAL segments 4), a behaviour we found for the transient {\em pulsed} component for energies above $\sim 20$ keV. Apparently, the increase of the {\em pulsed} emission above 20 keV is delayed with respect to that of the unpulsed-emission above 20 keV.
The contribution of the pulsed component to the total flux (=pulsed fraction) in the 20--150 keV band during INTEGRAL segment 4 amounts $34\pm 9\%$. At the start of the Jan. 2009 outburst this fraction was $31\pm7\%$, and about 240 days later during INTEGRAL segment 5 $45\pm 14\%$.


\section{Summary}
\label{sec:sum}
In this paper we presented new and detailed characteristics of the persistent (=non-burst) emission
of \axpeinstein\ between $\sim 1$ and 300 keV analyzing data collected over a period of
27 months, starting with the onset of an outburst in October 2008, covering the epoch of the
January 2009 outburst, and ending two years later in January 2011. Detailed results are
derived on the timing characteristics, the evolutions of the total and pulsed persistent 
emissions, and pulse profiles. In particular, the evolution of the total and pulsed emission
of a magnetar for energies above $\sim 10$ keV after the event of a major timing glitch accompanied 
with a radiative outburst has been revealed for the first time. We used RXTE PCA and Swift XRT monitoring 
data to derive incoherent and coherent timing models for the pulse frequency evolution between October 2008 
and January 2011. RXTE PCA and HEXTE data were used to derive the evolution of the pulse profile, flux and 
spectrum of the pulsed emission from $\sim 2.5$ to 150 keV. Finally, we exploited the imaging capabilities 
of Swift XRT and INTEGRAL ISGRI to derive the evolution of the flux and spectra of the {\em total\/} emission 
for energies from $\sim 1$ to 300 keV after the January 2009 outburst. What follows is a summary of the main 
results.


\subsection{Total flux evolution of \axpeinstein\ below 10 keV}

1) Using RXTE-PCA data we derived the total flux evolution for energies $\sim 2 - 10$ keV from Oct. 3, 2008 up to Dec. 25, 2010 showing the outburst in October 2008 and the much more intense outburst in January 2009. The flux decay of the latter has still not reached its quiescent level $\sim$2 years
after the outburst (\textit{Sect.\,\ref{rxte_evol_total}, upper panel of Fig.\,\ref{evolution_flux_nu_pflux}}). See also \citet{israel2010, bernardini2011, scholz2011}.

2) We can satisfactorily describe the decay of the total flux after the intense Jan. 22, 2009 (MJD 54853.911) outburst with a rapidly decaying power-law component, dominating only during a period of $\sim$5 days after the outburst, and a slowly decaying exponential component with a typical time scale of $\sim$200 days (\textit{Sect.\,\ref{rxte_evol_total}}).

3) In the tail of the intense Jan. 22, 2009 outburst we detected two mini-outbursts, at MJD 54920 and at MJD 55207, with enhanced emission over periods of $\sim$450 s and $\sim$130 s, respectively, and superposed with short duration bursts. In the tail of the latter outburst we found in a time window of only 52 s a pulsed signal (energies $\sim$2 - 20 keV) of $\sim$13 $\sigma$ significance (\textit{Sect.\,\ref{sect:mini_outbursts}}). The latter event might be similar to the persistent, pulsed emission enhancement reported by \citet{kaneko10} using Fermi GBM for a time interval just after the onset of the Jan. 2009 outburst of \axpeinstein.

 
\subsection{Evolution of the pulse frequency of \axpeinstein\ and glitch detections}

1) From a timing analysis of RXTE PCA sub-observations we derived the evolution of the pulse frequency from Oct. 3, 2008 till Dec. 25, 2010, based on incoherent measurements (\textit{Sect.\,\ref{evolution_pulse_frequencies}, 
middle panel of Fig.\,\ref{evolution_flux_nu_pflux}}).

2) Two apparent discontinuities in the pulse-frequency evolution are visible, near MJD 54853
(the onset of the Jan. 22, 2009 outburst) and near MJD 55121 (Oct. 17, 2009). For three time intervals, before, in between 
and after the epochs of these timing discontinuities, we derived simple incoherent timing models
(Taylor series) with either 3 or 4 timing parameters ($\nu,\dot\nu,\ddot\nu[,\dddot\nu]$) (\textit{Sect.\,\ref{incoherent_models},
Table\,\ref{ineph_table}}).

3) Phase-coherent timing models have been constructed for 6 time intervals, each of duration $\sim$30 - 35 days, using RXTE PCA and Swift XRT overlapping observations. Most importantly, one phase-connected ephemeris was derived after the onset of the Oct. 2008 outburst (see also \citet{ng2011} and \citet{israel2010}, and one just before the January 2009 outburst and two directly after this second outburst (\textit{Sect.\,\ref{coherent_models} and Table\,\ref{eph_table}}).

4) Applying standard glitch fitting techniques we measured at the time of the onset of the Jan. 2009 outburst (i.e. MJD 54853.035) a frequency jump $\Delta\nu/\nu$ of $(1.9\pm1.6)\cdot 10^{-6}$, and frequency derivative jump $\Delta\dot{\nu}/\dot{\nu}$ of $-0.69\pm0.07$. The frequency derivative jump value of $\Delta{\dot\nu}=+(1.30\pm0.14)\cdot 10^{-11}$ Hz/s indicates a dramatic sudden decrease of the spin-down rate, not seen before from any magnetar (\textit{Sect.\,\ref{glitchjan09}}; see also discussion section, Sect.\,\ref{sect_disc}).

5) For the discontinuity in the timing behaviour of \axpeinstein\ near MJD 55121 (Oct. 17, 2009), we estimated from the incoherent timing models that a strong $\Delta\dot{\nu}/\dot{\nu}$ jump occurred of size ($-0.89\pm 0.19$), similar to the Jan. 22, 2009 glitch. Note, this glitch was not accompanied by a radiative outburst (\textit{Sect.\,\ref{glitchoct09}}).
 
 
\subsection{Pulsed flux evolution of \axpeinstein\ below 10 keV}

1) Applying our derived timing models for phase-folding barycentered RXTE PCA event arrival times, we derived pulse-phase distributions for every PCA sub-observation. This allowed us to derive the pulsed flux evolution for energies $\sim 2 - 10$ keV from Oct. 3, 2008 up to Dec. 25, 2010 (\textit{Sect.\,\ref{pulsed_flux} and bottom panel of Fig.\,\ref{evolution_flux_nu_pflux}}). \citet{scholz2011} showed a consistent evolution covering the first $\sim 150$ days, including the two outburst epochs.

2) The evolution of the pulsed flux (energies $\sim 2 - 10$ keV) behaves very different compared to that of the total flux below 10 keV: the radiative outburst of the pulsed flux in October 2008 is stronger than that in January 2009, contrary to the findings for the total flux (\textit{Sect.\,\ref{pulsed_flux} and Fig.\,\ref{evolution_flux_nu_pflux}}). 

3) In October 2008 the pulsed flux ($\sim 2 - 10$ keV) reaches its maximum $\sim 10$ days after the start of the outburst of the total emission (\textit{bottom panel of Fig.\,\ref{evolution_flux_nu_pflux}}). See also \citet{ng2011, scholz2011}.

4) After the start of the January 2009 outburst the pulsed flux ($\sim 2 - 10$ keV) was unstable, varying by a factor of $\sim 2$ for two weeks and then a linear decay set in over a period of almost 700 days with a decay rate of $(-6.15 \pm 0.66)\times 10^{-4}$ c/(s day) (\textit{Sect.\,\ref{pulsed_flux} and bottom panel of Fig.\,\ref{evolution_flux_nu_pflux}}).


\subsection{Evolution of the pulse-profile morphology below $\sim 30$ keV}

1) Using RXTE PCA monitoring data we constructed for 11 time segments between Oct. 3, 2008 and Dec. 25, 2010 pulse profiles in three X-ray bands: $< 4.14$ keV (soft band), $4.14-11.22$ keV (medium band) and $11.22-33.82$ keV (hard band) (\textit{Sect.\,\ref{xray_pp_morph} and Fig.\,\ref{morph_vs_e_t}}). 

2) Between the onset of the October 2008 outburst and that of January 2009, in the soft and medium bands a single broad pulse is detected with a peak at pulse phase $\sim$ 0.9. In the same time window first evidence was found for weak pulsed emission in the hard band above 11 keV (\textit{Sect.\,\ref{xray_pp_morph} and Fig.\,\ref{morph_vs_e_t}}).

3) Drastic changes in pulse shape are found as a function of energy and of time after the glitch epoch and start of the radiative outburst in January 2009: most notably, a transient hard X-ray pulse appears in the phase distribution after $\sim$ 11 days, that can be well described with a Gaussian shape, centered at phase $\sim 0.59$ and width (FWHM) 0.23. In the hard band above $\sim 11$ keV this high-energy pulse is the single pulse detected; in the medium band this high-energy pulse is seen in addition to the broad pulse detected in the soft band, which is similar but not identical to that detected in the soft and medium bands before the January 2009 outburst (\textit{Sect.\,\ref{xray_pp_hard}, Fig.\,\ref{morph_vs_e_t}, Sect.\,\ref{xray_pp_medium}, Fig.\,\ref{hard_pflux_evol}\/}).

4) The new high-energy pulse reaches maximal flux in a time interval between 15 and 102 days after the onset of the January 2009 outburst and smoothly fades to undetectable levels within $\sim$ 350 days from the onset (\textit{Fig.\,\ref{hard_pflux_evol}\/}).

5) The time interval of $\sim 11$ days after the glitch in January 2009 appears to be a transition period in which the pulse profile changes from the single broad soft pulse before the glitch to the structured pulse consisting of a soft and a transient hard component with peak phase separation $\sim$ 0.3. This $\sim$ 11-day time interval is also the period in which the pulsed flux ($\sim 2 - 10$ keV) was unstable, varying by a factor of $\sim 2$ before a linear decay set in over a period of almost 700 days (\textit{Fig.\,\ref{morph_vs_e_t} and for variability Sect.\,\ref{pulsed_flux} and bottom panel of Fig.\,\ref{evolution_flux_nu_pflux}\/}).


\subsection{Evolution of the pulsed spectra for X-ray energies 2.5 - 150 keV}

1) Exploiting the simultaneous monitoring with the PCA and HEXTE aboard RXTE we derived the spectra of the pulsed emission over the broad energy range 2.5 - 150 keV in 11 time segments between Oct. 3, 2008 and Dec. 25, 2010 (\textit{Sect.\,\ref{X_total-pulsed_spectra}
and Fig.\,\ref{pulspc_evol}\/}).

2) Below $\sim$ 10 keV all pulsed-emission spectra are very soft with photon indices $\Gamma$ in the range -[3.9 - 4.2], except in the first time segment (11 days; RXTE segment-3) after the Jan. 2009 outburst ($\Gamma \sim -4.6$). In the time period between the two outbursts, the flux (normalization) was highest during the first segment (26 days), decreased by a factor of $\sim 2$ in the second segment (75 days) before the outburst in Jan. 2009. Promptly after the latter outburst, over a period of 11 days, the normalization was again higher by a factor $\sim$ 2, followed by a gradual decrease by a factor $\sim 2$ over almost 700 days (\textit{Sect.\,\ref{X_total-pulsed_spectra} and Fig.\,\ref{pulspc_evol}}).

3) Around $\sim$ 10 keV the pulsed spectra change from very soft spectra to hard non-thermal spectra, with jumps in spectral photon index in the range 2.5 - 3.0 (\textit{Fig.\,\ref{pulspc_evol}}), similar to what has been reported for the persistent pulsed emission of some AXPs \citep[e.g.][]{kuiper06}.

4) Above $\sim 10$ keV the variations are more drastic than at lower energies. Between the Oct. 2008 and Jan. 2009 outbursts the pulsed emission above 10 keV was too weak to derive a spectrum. Directly after the Jan. 2009 glitch/outburst, during the 11 days of segment 3, luminous pulsed
hard X-ray emission was detected up to $\sim 150$ keV. In the next time segment (\#4) the maximal flux was reached at $70\pm 30$ days after the glitch epoch, and a gradual decrease by more than a factor 10 followed over a period of $\sim 300$ days till the pulsed hard X-ray emission became undetectable. This spectral evolution follows the appearance and gradual fading of the hard X-ray pulse centered at phase $\sim 0.59$ in the pulse profile (\textit{Sects.\,\ref{xray_pp_hard}, \ref{X_total-pulsed_spectra} and Figs.\,\ref{hard_pflux_evol}, \ref{pulspc_evol}\/}).

5) During the periods with significant pulsed hard X-ray emissions, the total pulsed spectra over the broad energy range 2.5 - 150 keV can be satisfactorily described by a combination of a soft power-law (dominating below 10 keV) plus a cutoff power-law model (describing the hard X-ray emission). The latter component of the spectral fits reaches maxima in luminosity (in $\nu F_{\nu}$) for energies in the range 90 - 105 keV and the power-law photon index varies with energy from $\sim 0$ at 30 keV, $\sim -1.2$ at 70 keV to $\sim -2$ around 90 keV (\textit{Sect.\,\ref{X_total-pulsed_spectra}}).

The actual extrapolation of the spectral shape to higher energies (sharp break, bend?) is not constrained by our data.


\subsection{Evolution of the total spectra for X-ray energies 1 - 300 keV}

1) Making use of the imaging capabilities of INTEGRAL ISGRI, JEM-X and Swift XRT, we derived the spectral evolution of the total emission of \axpeinstein\ over the broad energy range 1--300 keV starting 2 days after the Jan. 2009 glitch/outburst till April 2010
(\textit{Sects.\,\ref{swift_xrt_tot}, \ref{integraltotsp} and Fig.\,\ref{totspc_evol}\/}).

2) The total spectra can be satisfactorily described by a black-body plus a single power-law model over the full 1-300 keV band. The black-body component was found to have $kT$ values varying in the range 0.57--0.74 keV. The photon index of the power-law component exhibited a hardening with time and a clear correlation with flux in the 20--150 keV band, namely, its value gradually changed from $\sim -1.4$ directly after the glitch in Jan. 2009 to $\sim -0.9$ in Febr.-April 2010, while the 20--150 keV flux decreased by a factor $\sim 3$ (\textit{Sect.\,\ref{integralswifttotsp}, Table\,\ref{totspec_prop} and Fig.\,\ref{totspc_evol}\/}).

3) The total soft X-ray flux (1--10 keV) measured with Swift XRT (contributions from the black-body plus power-law components) is maximal during the first INTEGRAL observation two days after the glitch and decays similarly to what we measured for the count rates below 10 keV with RXTE PCA (\textit{Sect.\,\ref{integralswifttotsp}, Figs.\,\ref{totspc_evol} and \ref{totflux_evol}\/}).
 
4) The total hard X-ray flux (20--150 keV) measured with INTEGRAL ISGRI is also directly maximal in the first observation 2 days after the glitch in Jan. 2009, decaying in flux by a factor $\sim$ 3 till Febr.-April 2010, when the source was still detected in the sky map at a 9.5$\sigma$ level in the 20--150 keV band (\textit{Sect.\,\ref{integraltotsp}, Figs.\,\ref{isgri_map}, \ref{totspc_evol} and \ref{totflux_evol}\/}). This behaviour is in contrast to what we measured for the evolution of the pulsed hard X-ray component: the latter reached maximal flux $70\pm 30$ days after the glitch, subsequently decaying to undetectable levels in $\sim 300$ days, a decrease in flux by more than a factor of 10 (\textit{Figs.\,\ref{hard_pflux_evol} and \ref{pulspc_evol}\/})

\section{Discussion}

\label{sect_disc}

The above summarized results on the evolution of the high-energy characteristics of \axpeinstein\ over a period of 27 months,
adressing the timing parameters, pulse morphologies, total and pulsed spectra in the soft and hard X-ray band, provide important constraints on the theoretical modelling of the production scenario's in the magnetosphere. Particularly interesting is that for the first time from the start of an outburst, accompanied by a timing glitch, not only the evolution at soft X-ray energies could be studied in detail, but also at (non-thermal) hard X-ray energies, and the behaviours of distinctly different components compared. The measured timing glitch in January 2009 is itself already interesting, so far being the most extreme {\it instantaneous} frequency derivative jump $\Delta\dot{\nu}/\dot{\nu}$ of $-0.69\pm0.07$ detected, with $\Delta{\dot\nu}=+(1.30\pm0.14)\cdot 10^{-11}$ Hz/s. 

We scrutinized earlier publications on magnetar (SGR/AXP) timing, mainly based on RXTE observations, and focussed on frequency derivative jumps $\Delta\dot{\nu}/\dot{\nu}$ with {\it negative} signs of order unity. Although large gradual $\dot{\nu}$ fluctuations are reported for SGR 1900+14 \citep{woods1999,woods2002}, and SGR 1806-20 \citep{woods2002,woods2007}, no convincing {\it sudden} instantaneous $\dot{\nu}$ jumps are detected. A possibly similar event as observed for \axpeinstein\ may have occured for SGR 1900+14 \citep{woods2002} near its Aug. 27, 1998 giant flare of size $\Delta{\dot\nu}/\dot{\nu}\sim -0.28$, but due to a $\sim 80$ days data gap in the RXTE timing observations prior to the giant flare a gradual accelerated spin-down could not be excluded.
A radiative (pulsed and persistent emission) outburst, accompanied with a timing glitch, has been reported by \citet{kaspi03} for AXP 1E 2259+586, but the measured $\Delta{\dot\nu}/\dot{\nu}$ jump of $+1.11$ has positive sign, and the absolute value of the $\Delta{\dot\nu}$ jump
is orders of magnitude smaller than that observed for \axpeinstein. Also, the glitches with a measurable $\Delta{\dot\nu}$ jump detected from AXPs 1RXS J1708-40 \citep{dib08}, 1E 1841-045 \citep{dib08}, 4U 0142+614 \citep{dib07,gavriil11} and 1E 1048.1-5937 \citep{dib09} do not show the magnitude, sign and relative strength as observed for \axpeinstein, making the latter glitch unique.

The dramatic decrease in spin-down rate, observed for \axpeinstein\, during the Jan. 22, 2009 glitch, would for purely dipole spin-down naively imply a decrease in effective surface dipole field strength by  $\sim 50\%$. In the twisted magnetosphere model such a drastic change in $\dot{\nu}$ and B might be due to a varying twist angle with corresponding spectral and flux changes \citep{beloborodov2007}. 
The epoch of the timing glitch is consistent with the start of the outburst, therefore, we assume that the glitch triggered the outburst on MJD 54853.035 (TDB), Jan. 22, 2009. It is then intruiging that there are no signs of a radiative outburst near MJD 55121, Oct. 17, 2009, for which our analysis suggests a similarly strong $\Delta\dot{\nu}/\dot{\nu}$ jump.

Recently published works on \axpeinstein\, of which the analyses were performed in parallel to our work, focussed primarily on the evolution of the (pulsed) emission below 10 keV in time intervals up to $\sim$ 20 days after the outbursts in Oct. 2008 and Jan. 2009 \citep{ng2011, bernardini2011, scholz2011}, as we summarized in the introduction. 

In our work, we revealed in addition the creation of a new transient non-thermal pulse in the pulse phase distribution after the Jan. 2009 glitch, and could study the simultaneous evolution of the pulsed and total high-energy emissions, the latter being dominated by an unpulsed component ($\sim 65\%$). It is striking, that the spectrum of the pulsed emission from 2 keV up to 150 keV after the glitch is very similar to the shapes of the broad-band spectra of the persistent pulsed emissions of three AXPs, 1E 1841-045, 4U 0142+61 and 1RXS J1708-40 \citep{kuiper06,denhartog2008a,denhartog2008b}. The non-thermal pulsed component (20--150 keV) of \axpeinstein\ reaches in the time interval 15--101 days after the glitch (RXTE time segment-4; see Table \ref{rxte_segments}) a flux of $(7.5^{+0.9}_{-1.0})\cdot 10^{-11}$ erg/cm$^2$s (see Table \ref{pulsed_flux_tab}) and an (isotropic) luminosity of $\sim 13.6\cdot 10^{34}$ erg/s adopting the distance estimate of 3.9 kpc of \citet{tiengo10}, which is about 1.4 times the spin-down luminosity during that time segment. There are clearly two maxima in the luminosity distribution of the pulsed spectrum, one around 100 keV and one near 1 keV (the maximum, likely due to a BB component, is reached below our energy interval; earlier works did not address the spectra of the {\sl pulsed} emission of \axpeinstein\ below 10 keV, nor above 10 keV). We also see a clear phase separation of $\sim$ 0.3 between the soft-spectrum pulse and the hard non-thermal pulse, similar to the case of 1RXS J1708-40 \citep[e.g.][]{denhartog2008b}. In fact, also the broad-band total spectrum is reminiscent of those measured for these three persistent AXPs. However, for \axpeinstein\ we are dealing with a transient / variable phenomenon above (and below) 10 keV, while it was found that the non-thermal persistent emission of AXPs appeared stable within the statistical errors of $\sim$ 20\% over more than 10 years. 

The steadiness of the persistent non-thermal emission from AXPs led \citet{beloborodov2009, beloborodov2011} to reconsider his scenario's proposed earlier to explain the production of this non-thermal component \citep{beloborodov2007}. As mentioned in the introduction, he discusses how a starquake  can cause convective motions in the crust which twist the external magnetic field anchored to the surface. The magnetic twist energy is dissipated over time in the form of radiation, but the untwisting occurs in a peculiar way leading to the creation of a bundle of electric currents, the so-called j-bundle, which shrinks toward the magnetic dipole axis. The high-energy component, created by upscattering of thermal X-rays from the neutron star surface by an inner relativistic outflow in the quasi-steady j-bundle, is beamed along the magnetic dipole axis. The latter can explain the relatively narrow high-energy pulses. Furthermore, his radiative-transfer simulations produce spectral shapes that get close to those observed, but with possibly too high break energies near 1 MeV.

It is very interesting to note that a radiative outburst accompanied by a strong glitch in the case of \axpeinstein\ indeed led to a scenario/geometry which mimics in many respects the observational characteristics of the persistent emission seen from AXPs. But our results on \axpeinstein\ pose several new observational constraints on this model scenario, as well as on other theoretical attempts (none of the other models addresses the transient hard X-ray ($>20$ keV) phenomenon of the persistent emission). A constraint on the time scale of creation of the quasi-steady narrow j-bundle, is the delay with respect to the glitch epoch of $70\pm 30$ days, with which the high-energy pulse with non-thermal spectrum was created. Another important finding is that the total non-thermal high-energy emission
was present immediatly in the first INTEGRAL observation starting 2 days after the glitch. This suggests that there should be a ``corona'' around the  neutron star in which non-thermal unpulsed emission is produced without delay already at the start of the outburst. The model scenario of \citet{beloborodov2009, beloborodov2011} was proposed to explain the steady non-thermal emission from AXPs. What we find is a transient pulsed component which decays by a factor $\ga 10$, becoming undetectable $\sim 300$ days after the glitch. However, the 
non-thermal total (pulsed+unpulsed) emission decayed only by a factor $\sim 3$ over more than a year, seemingly stable over the last $\sim$ 100 days. A possibility is that a steady state similar to that of persistent emission of the above mentioned AXPs was reached after about one year, but that the pulsed non-thermal emission
has become too weak to be detectable with currently operational high-energy instruments. The situation might be representative for all those AXPs for which no non-thermal emission has been detected (yet), but all have X-ray fluxes in the transition region around 10 keV that also are weaker than detected for the AXPs with reported steady non-thermal persistent emission above 20 keV up to $\sim 150$ keV. In this respect, we like to note that \axpeinstein\ was in the field of view of INTEGRAL during revolutions 470--572 (Aug. 21, 2006 -- June 22, 2007) for an effective exposure of 227.2 ks during a period in which the soft X-ray flux state increased at least $16 \times$ its historical flux minimum reached in July--August 2006. \citet{halpern08} concluded that the source was then recovering from an X-ray outburst in between August 2006 and June 2007. However, we did not find evidence for an outburst of non-thermal emission above 20 keV from \axpeinstein\ in the INTEGRAL ISGRI sky maps. The $2\sigma$ upper limits to the flux measurements are at the level of the positive flux measurements in Revs, 899--912 (around March 2010) in Fig.\,\ref{totspc_evol}, but still significantly higher than the low flux upper limits from the very deep summed exposure before 2006 (Revs. 46--411 in Fig.\,\ref{totspc_evol}). Also in this case we cannot exclude the scenario proposed by \citet{beloborodov2009, beloborodov2011}, because the total and particularly pulsed emission from AXPs and also from SGRs is (too) weak for easy detection by presently operational instruments.

We noted the importance of studying the characteristics of magnetars over a broad energy interval. In the case of \axpeinstein\ we concluded that discusssions based on spectral model fits to the narrow energy range 1--10 keV \citep[e.g.][]{scholz2011} leads to apparent characteristics (e.g. power-law component with spectral softening with decreasing flux) which have to be revised when the broad
energy range is considered, including the hard X-rays (power-law component with spectral hardening with decreasing flux). Furthermore, broad-band studies of the magnetar class are now becoming possible, as has been published by \citet{enoto2010b} using Suzaku observations of a sample of eight magnetars, including \axpeinstein\ in outburst. In that paper interesting correlations have been studied for the total emission of the hardness ratio, defined as the ratio of the flux of the hard, non-thermal component (power-law fit; 1--60 keV) and the flux of the soft component (a black body, a comptonized black body or a combination of the two; 1--60 keV) with derived characteristics as e.g. characteristic age and magnetic field strength. Such a study ignores, however, that the origins of the actual unpulsed and pulsed emissions appear to be different (location in the magnetosphere, production scenario) as is evidenced by the different spectra of the pulsed and total (pulsed+unpulsed) emissions (particularly above 10 keV), the different pulse profiles (shape and phase) for soft and hard pulsed  components, and the different evolutions of all contributing components. Theoretical modelling of the physics taking place under the extreme conditions in magnetospheres of magnetars is extremely complex. It is encouraging that the amount and detail of observational constraints is increasing significantly over the last years.

\acknowledgements
We acknowledge the use of public data from the Swift data archive.
This research has made use of data obtained from the High Energy Astrophysics 
Science Archive Research Center (HEASARC), provided by NASA's Goddard Space Flight Center,
and of data obtained through the INTEGRAL Science Data Centre (ISDC), Versoix, Switzerland.
INTEGRAL is an ESA project with instruments and science data centre funded by ESA member 
states (especially the PI countries: Denmark, France, Germany, Italy, Switzerland, Spain), 
Czech Republic and Poland, and with the participation of Russia and the USA.
We have extensively used NASA's Astrophysics Data System (ADS).
This work has been supported by NASA grant NNX10AJ54G.

\clearpage 

\end{document}